\documentclass[twocolumn,english,pra,showpacs,superscriptaddress]{revtex4}
\usepackage[T1]{fontenc}
\usepackage[latin9]{inputenc}
\usepackage{color}
\usepackage{bm}
\usepackage{amsmath}
\usepackage{amssymb}
\usepackage{graphicx}
\usepackage{esint}

\makeatletter

\@ifundefined{textcolor}{}
{%
 \definecolor{BLACK}{gray}{0}
 \definecolor{WHITE}{gray}{1}
 \definecolor{RED}{rgb}{1,0,0}
 \definecolor{GREEN}{rgb}{0,1,0}
 \definecolor{BLUE}{rgb}{0,0,1}
 \definecolor{CYAN}{cmyk}{1,0,0,0}
 \definecolor{MAGENTA}{cmyk}{0,1,0,0}
 \definecolor{YELLOW}{cmyk}{0,0,1,0}
}

\usepackage[colorlinks,linkcolor=blue,anchorcolor=blue,citecolor=blue,urlcolor=blue]{hyperref}

\makeatother

\usepackage{babel}
\begin{document}

\title{Polaronic effects in one- and two-band quantum systems}

\author{Tao Yin}

\email{taoyin@itp.uni-frankfurt.de}

\selectlanguage{english}%

\address{Institut f\"{u}r Theoretische Physik, Goethe-Universit\"{a}t, 60438
Frankfurt/Main, Germany}

\author{Daniel Cocks}

\address{College of Science, Technology \& Engineering, James Cook University,
Townsville 4810, Australia}

\author{Walter Hofstetter}

\address{Institut f\"{u}r Theoretische Physik, Goethe-Universit\"{a}t, 60438
Frankfurt/Main, Germany}

\date{\today}
\begin{abstract}
In this work we study the formation and dynamics of polarons in a
system with a few impurities in a lattice immersed in a Bose-Einstein
condensate (BEC). This system has been experimentally realized using
ultracold atoms and optical lattices. Here, we consider a two-band
model for the impurity atoms, along with a Bogoliubov approximation
for the BEC, with phonons coupled to impurities via both intraband
and interband transitions. We decouple this Fr\"{o}hlich-type term
by an extended two-band Lang-Firsov polaron transformation using a
variational method. The new effective Hamiltonian with two (polaron)
bands differs from the original Hamiltonian by modified coherent transport,
polaron energy shifts, and induced long-range interaction. A Lindblad
master-equation approach is used to take into account residual incoherent
coupling between polaron and bath. This polaronic treatment yields
a renormalized inter-band relaxation rate compared to Fermi's golden
rule. For a strongly coupled two-band Fr\"{o}hlich Hamiltonian, the
polaron is tightly dressed in each band and can not tunnel between
them, leading to an \textit{inter-band self-trapping} effect. 
\end{abstract}


\maketitle

\section{Introduction\label{sec:Introduction}}

The field of ultracold atom physics has explored a wide variety of
phenomena since its relatively recent accessibility, with a major
feature being the tunability of experiments across wide parameter
regimes \cite{Feynman1982,Lloyd1996,Jaksch1998,Cirac2012}, to easily
access and probe phase transitions \cite{Bloch2008,Bloch2012}, as
well as excitation spectra and dynamics of systems analog to condensed
matter \cite{Gericke2008,Bakr2009,Sherson2010,Serwane2011}. Even
features such as artificial gauge fields can be implemented for neutral
atomic particles, allowing for the investigation of topological phases
\cite{Lin2009,Cocks2012}. 

Within this ultracold toolbox, one ingredient is becoming of increasing
interest in recent years which is of vital importance to real solid-state
systems: phonons and atom-phonon coupling \cite{Herrera2011,Mostame2012,Hague2012,Hague2012a}.
Such a coupling provides many interesting possibilities \cite{Mathey2004,Das1999,Marchand2010}.
For one, it can lead to effective Hamiltonians, such as extended Hubbard
models or the Holstein model \cite{Holstein1959,Mahan2000,Hohenadler2004,Golez2012,Marchand2013},
as well as dissipative two-level system \cite{Zwerger1983,Leggett1987,Agarwal2013,Bera2014a}.
Polaronic effects from electron-phonon interactions have also long
been suggested to be the proponent behind high-$T_{c}$ superconductivity
in one- and two-band solid-state systems \cite{Zaanen1988,Gammel1990,Furukawa1994,Lanzara2001,Devreese2009,Alexandrov2010,Kagan2012}.
In ultracold quantum gases, evidence of polarons has been found in
systems with trapped ions \cite{Stojanovic2012} or systems with a
single ion immersed in a degenerate quantum gas \cite{Grier2009,Zipkes2010,Schmid2010,Rellergert2011,Ratschbacher2012,Ratschbacher2013,Casanova2012}.
On the other hand, the atomic polaron has also been studied both experimentally
and theoretically in systems of imbalanced Bose-Fermi mixtures \cite{Gunter2006,Ospelkaus2006,Best2009,Will2011,Leskinen2010,Casteels2012}
and Fermi-Fermi mixtures \cite{Prokofev2008,Mora2009,Schirotzek2009,Nascimbene2009,Yi2012b,Massignan2014,Cetina2015,Lan2015}.
In the particular case of a system with impurities immersed in a bosonic
bath, these impurities couple to bosonic excitations. For suitable
parameters, polaronic phenomena arise generically in such systems
\cite{Klein2005,Cucchietti2006,Bruderer2007,Bruderer2008,Tempere2009,Gadway2010,Privitera2010,Fukuhara2013,Casteels2013a,Dutta2013,Rath2013,Benjamin2014,Compagno2014,Li2014,Shashi2014,Grusdt2014,Grusdt2014a,Shchadilova2014,Christensen2015a,Levinsen2015,Schmidt2015,Hohmann2015,Ardila2015,Grusdt2015a}.
There are also other proposals for realizing atom-phonon couplings
and polaronic effects, including crystals of dipolar molecules \cite{Pupillo2008,Ortner2009,Herrera2013},
nanoparticles \cite{Gullans2012,Mottl2012} and hybrid atom-ion coupled
systems \cite{Bissbort2013}. 

None of these works consider polaronic phenomena of multi-band systems
with ultracold quantum gases. In such systems, the inter-band dynamics
of polarons, as well as intra-band dynamics, lead to new effects which
have no analogue in single band systems. Motivated by\textcolor{red}{{}
}recent experiments \cite{Heinze2011,Spethmann2012,Scelle2013,Chen2014},
we consider a system of a few impurities in an optical lattice, populating
the lowest two Bloch bands, immersed in a Bose-Einstein condensate.
These impurities are coupled to Bogoliubov phonons (of the BEC) via
both intra- and inter-band transitions. In order to decouple this
Fr\"{o}hlich-like term, we derive a generalized two-band Lang-Firsov
polaron transformation. The transformed effective Hamiltonian still
contains two bands, where the impurity is now dressed by phonons as
a quasi-particle (polaron). We use a variational approach to connect
between the weak and strong coupling limits and calculate the dressing
parameters. Polaronic effects modify both intra-band coherent transport
and polaron energy shifts, and also induce a long-range interaction
between different polarons. 

We then focus on inter-band relaxation effects and specify our system
as a single impurity trapped in a quasi-1D system. We study the residual
incoherent coupling between polaron and bath by using a Lindblad master
equation. The impurity inter-band relaxation process under this polaronic
treatment is beyond a Fermi's Golden Rule description. These polaronic
renormalization effects of the inter-band relaxation rate should be
accessible in current experiments. On the other hand, for large impurity-phonon
coupling, the polaron is tightly dressed in each band and cannot hop
between different bands. In this limit, an \textit{inter-band self-trapping}
effect is expected. 

This work is organized as follow: in section \ref{sec:Effective-Hamiltonian},
we introduce the effective two-band Hamiltonian of a realistic experiment
setup, with a few impurities in a lattice, immersed in a Bose-Einstein
condensate (BEC) of a different atomic species. In section \ref{sec:Variational-Polaron-transformati},
we describe in detail the generalized Lang-Firsov polaron transformation
for the two-band system. The transformed Hamiltonian with two (polaron)
bands can be separated into a coherent part and an incoherent part.
The coherent part, arising from the thermal average over the phonon
bath, is discussed in section \ref{sec:coherent-part}. The inter-band
relaxation and decoherence effects, which are all included by the
incoherent part of this Hamiltonian, are discussed in section \ref{sec:Lindblad-equation}.
We derive the Lindblad equation and correlation functions for the
residual incoherent impurity-phonon coupling. This polaronic impurity
dynamics is closely related to recent experiments. The polaronic inter-band
relaxation rate is compared to Fermi's Golden Rule. We give concluding
remarks and an outlook in section \ref{sec:Conclusion}.

\section{Effective Hamiltonian\label{sec:Effective-Hamiltonian}}

Here we consider a few neutral impurities with mass $m_{I}$ interacting
with a Bose-Einstein condensate of another neutral species. The impurities
are trapped by a 3D optical lattice and their Hamiltonian is denoted
by $H_{I}$. The homogeneous BEC system $H_{B}$ is formed by another
atomic species with mass $m_{B}$ and a weak repulsive interaction
$g_{B}$. The impurity-BEC interaction $H_{\text{int}}$ is caused
by $s$-wave interactions between the different species, which can
be tuned by standard Feshbach resonance techniques. The total Hamiltonian
is hence 
\[
H=H_{I}+H_{B}+H_{\text{int}}.
\]
We describe these different terms in detail in the following parts
of this section.

\subsection{Impurities in optical lattice -- $H_{I}$}

We consider a quasi-1D system with impurities trapped in an anisotropic
3D optical lattice:
\[
V_{I}\left(\mathbf{r}\right)=V_{I}^{x}\sin^{2}\left(\frac{\pi}{d}x\right)+V_{I}^{y}\sin^{2}\left(\frac{\pi}{d}y\right)+V_{I}^{z}\sin^{2}\left(\frac{\pi}{d}z\right),
\]
with lattice constant $d=\lambda/2$ and laser wavelength $\lambda$.
We use the lattice constant $d$ as the unit of length throughout
this paper. The single impurity recoil energy is $E_{R}\equiv\pi^{2}\hbar^{2}/\left(2m_{I}\right)$.
The trapping strength in the transverse ($y,z-$) direction is assumed
to be much stronger than in longitudinal ($x-$) direction with $V_{I}^{x}\ll V_{I}^{\perp}\equiv V_{I}^{y}=V_{I}^{z}$.
The impurities are therefore tightly trapped in the transverse direction,
and remain in the ground state of the associated harmonic oscillator
potential:
\begin{equation}
\phi^{0}\left(y\right)=1/\left(\pi\sigma_{\perp}^{2}\right)^{\frac{1}{4}}e^{-y^{2}/\left(2\sigma_{\perp}^{2}\right)},\label{eq:wannier_y}
\end{equation}
with transverse characteristic length $\sigma_{\perp}\equiv\sqrt{\hbar/\left(m_{I}\omega_{\perp}\right)}$
and frequency $\hbar\omega_{\perp}\equiv2\left(V_{I}^{\perp}E_{R}\right)^{1/2}$.
In the longitudinal direction, on the other hand, the impurities populate
both the lowest and the first excited band. In principle, exact Bloch
wave functions need to be calculated numerically, which can then also
be represented as linear combinations of Wannier states $\varphi_{x}^{0\left(1\right)}\left(x\right)$.
In a deep lattice, the Wannier functions can be approximated by the
lowest two eigenstates of a harmonic oscillator 
\begin{eqnarray}
\varphi_{x}^{0}\left(x\right) & \approx & 1/\left(\pi\sigma_{x}^{2}\right)^{\frac{1}{4}}e^{-x^{2}/\left(2\sigma_{x}^{2}\right)},\label{eq:wannier_x0}\\
\varphi_{x}^{1}\left(x\right) & \approx & 1/\left(\pi\sigma_{x}^{2}\right)^{\frac{1}{4}}\left(\sqrt{2}x/\sigma_{x}\right)e^{-x^{2}/\left(2\sigma_{x}^{2}\right)},\label{eq:wannier_x1}
\end{eqnarray}
with longitudinal characteristic length $\sigma_{x}\equiv\sqrt{\hbar/\left(m_{I}\omega_{x}\right)}$
and oscillation frequency $\hbar\omega_{x}\equiv2\left(V_{I}^{x}E_{R}\right)^{1/2}$. 

In this highly anisotropic system, impurities can hop to their nearest-neighbor
sites only along the longitudinal direction. A particle localized
on lattice site $j$ will be described by the wave function
\begin{equation}
W_{j}^{0\left(1\right)}\left(\mathbf{r}\right)\equiv\varphi_{x}^{0\left(1\right)}\left(x-x_{j}\right)\phi^{0}\left(y\right)\phi^{0}\left(z\right).
\end{equation}
Due to the low density of impurities, their dynamics can be modeled
by a non-interacting two-band Hubbard Hamiltonian:
\begin{alignat}{1}
\hat{H}_{I}= & -\sum_{\langle i,j\rangle}\sum_{\alpha}J^{\alpha}\hat{a}_{i}^{\alpha\dagger}\hat{a}_{j}^{\alpha}+\sum_{i}\sum_{\alpha}\varepsilon^{\alpha}\hat{n}_{i}^{\alpha}
\end{alignat}
where $J^{\alpha}$ and $\varepsilon^{\alpha}$ are the hopping parameters
and on-site energy for each band with index $\alpha=0,1$. In a deep
lattice, the band gap $\varepsilon^{\Delta}\equiv\varepsilon^{1}-\varepsilon^{0}$
can be approximated as the longitudinal oscillator frequency $\hbar\omega_{x}$.
In this work we only consider inter-band dynamics between the lowest
two Bloch bands and ignore higher band effects.

\begin{figure}
\includegraphics[bb=0bp 0bp 368bp 275bp,width=0.45\textwidth]{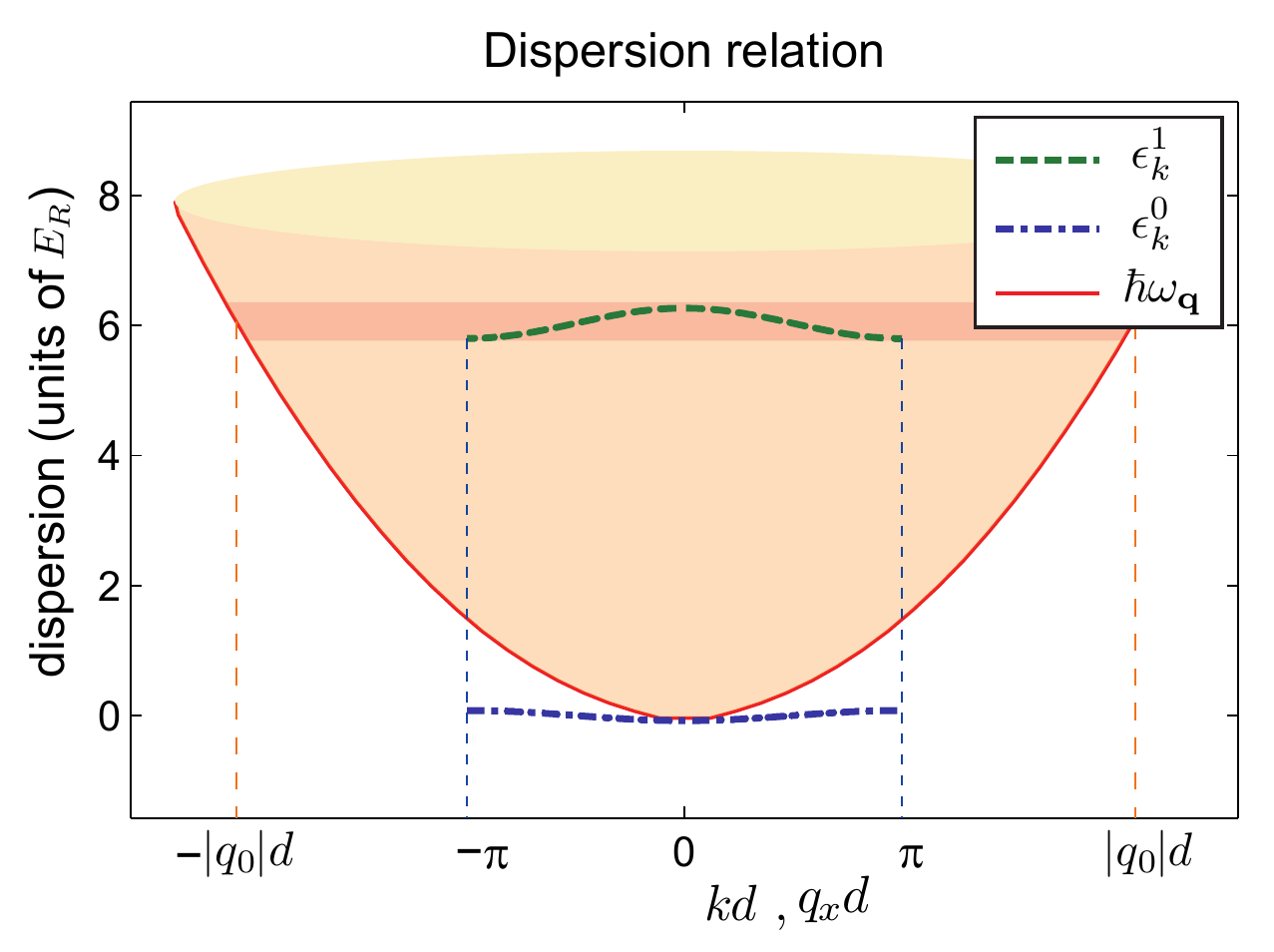}

\protect\caption{Dispersion relation for single impurity and phonon in longitudinal
direction (in the unit of $E_{R}$). The impurity moves in the longitudinal
direction of the system with two Bloch bands $\epsilon_{k}^{0}$ (blue
line) and $\epsilon_{k}^{1}$ (green line). The dashed blue lines
at $k=\pm\pi$ indicate the first Brillouin zone in longitudinal direction.
The phonon dispersion $\hbar\omega_{\mathbf{q}}$ (red line) is described
by Bogoliubov theory where $\mathbf{q}\equiv\left(q_{x},q_{y,}q_{z}\right)$
is the phonon momentum. The dashed red lines at $q_{x}=\pm|q_{0}|$
indicate that the phonon energy matches with the impurity band gap
$\hbar\omega_{|q_{0}|}\approx\varepsilon^{\Delta}$ with $\varepsilon^{\Delta}\equiv\varepsilon^{1}-\varepsilon^{0}$.\label{fig:dispersion}}
\end{figure}

\subsection{Bosonic bath}

The impurities are immersed in a homogeneous BEC with weakly repulsive
boson-boson interaction $g_{B}$ between the atoms. In a dilute system,
this weak interaction can be described by boson-boson scattering length
$a_{B}$ as $g_{B}=4\pi\hbar^{2}a_{B}/m_{B}$. For vanishing inter-species
interaction $g_{IB}$ between impurity and bath, the BEC can be described
by standard Bogoliubov theory and treated as a phonon bath (see Fig.
\ref{fig:dispersion}). Once $g_{IB}$ is introduced, the BEC becomes
deformed due to the presence of impurities. This interaction is closely
related to the impurity-boson scattering length and other system parameters
such as the impurity-boson mass ratio and the impurity confinement
strength. The relation can be determined by making use of scattering
theory in the low-energy limit, such as the Lippmann-Schwinger equation
or effective field theory \cite{Braaten2006}. For an unconfined impurity,
the inter-species interaction $g_{IB}$ can be derived as $g_{IB}\equiv2\pi\hbar^{2}a_{IB}/\mu$
with the reduced mass $\mu\equiv m_{I}m_{B}/\left(m_{I}+m_{B}\right)$
and 3D impurity-boson scattering length $a_{IB}$. On the other hand,
for the confined impurity, $g_{IB}$ needs to be treated carefully
due to lattice effects such as confinement induced resonances \cite{Olshanii1998,Massignan2006}.
In the specific system we considered here, the impurity is confined
in one-dimensional tube (quasi-1D) by an anisotropic optical lattice
while the bosonic atoms are free in three-dimensional space (3D).
When the transverse characteristic length $\sigma_{\perp}$ is much
smaller than any other length scales, the resulting system is mixed-dimensional.
At low energies, the inter-species interaction is solely characterized
by a single parameter, the effective scattering length $a_{IB}^{\text{eff}}$,
whose value can be obtained numerically \cite{Nishida2008}. This
fact also allow us to arbitrarily tune the value of $a_{IB}^{\text{eff}}$
by tuning the transverse confinement strength, independently of tuning
the Feshbach resonance position.

Here we use the approach in \cite{Bruderer2007,Bruderer2008}, where
the deformation is treated as a perturbation around the BEC ground
state. The bosonic field operator is expanded as $\hat{\psi}\left(\mathbf{r}\right)=\psi_{0}\left(\mathbf{r}\right)+\hat{\vartheta}\left(\mathbf{r}\right)$,
where $\psi_{0}\left(\mathbf{r}\right)$ is the order parameter in
the absence of inter species interaction and $\hat{\vartheta}\left(\mathbf{r}\right)=\vartheta\left(\mathbf{r}\right)+\hat{\zeta}\left(\mathbf{r}\right)$
represents the perturbation itself, which consists of a correction
to the order parameter, $\vartheta\left(\mathbf{r}\right)$, and Bogoliubov
excitation operators $\hat{\zeta}\left(\mathbf{r}\right)$. The modified
BEC ground state is described as $\psi_{0}\left(\mathbf{r}\right)+\vartheta\left(\mathbf{r}\right)$,
which is the Gross-Pitaevskii solution including the presence of impurities.
This modification $\vartheta\left(\mathbf{r}\right)$ shifts the equilibrium
positions in order to minimize the total energy of the system. The
small excitations $\hat{\zeta}\left(\mathbf{r}\right)$, around the
static GP ground state $\psi_{0}\left(\mathbf{r}\right)+\vartheta\left(\mathbf{r}\right)$
of the condensate, can be described in the terms of Bogoliubov modes
\begin{equation}
\hat{\zeta}\left(\mathbf{r}\right)=\sum_{\mathbf{q}}\left[u_{\mathbf{q}}e^{i\mathbf{q}\cdot\mathbf{r}}\hat{\beta}_{\mathbf{q}}-v_{\mathbf{q}}^{*}e^{-i\mathbf{q}\cdot\mathbf{r}}\hat{\beta}_{\mathbf{q}}^{\dagger}\right],\label{eq:phonon_beta}
\end{equation}
with bosonic operators $\hat{\beta}_{\mathbf{q}}^{\dagger}$ ($\hat{\beta}_{\mathbf{q}}$)
creating (annihilating) a Bogoliubov quasi-particle with momenta $\mathbf{q}$.
The coefficients $u_{\mathbf{q}}\left(\mathbf{r}\right)$ and $v_{\mathbf{q}}\left(\mathbf{r}\right)$
can be determined by Bogoliubov-de Gennes equations:
\begin{alignat*}{1}
u_{\mathbf{q}}= & \sqrt{1/\left(2\Omega\right)\left[\left(\epsilon_{\mathbf{q}}+g_{B}n_{0}\right)/\left(\hbar\omega_{\mathbf{q}}\right)+1\right]}\\
v_{\mathbf{q}}= & \sqrt{1/\left(2\Omega\right)\left[\left(\epsilon_{\mathbf{q}}+g_{B}n_{0}\right)/\left(\hbar\omega_{\mathbf{q}}\right)-1\right]}
\end{alignat*}
with the system \textit{quantization volume} $\Omega$ and the Bogoliubov
phonon dispersion
\begin{eqnarray}
\hbar\omega_{\mathbf{q}} & \equiv & \sqrt{\epsilon_{\mathbf{q}}\left(\epsilon_{\mathbf{q}}+2g_{B}n_{0}\right)};\nonumber \\
\epsilon_{\mathbf{q}} & \equiv & \frac{\hbar^{2}|\mathbf{q}|^{2}}{2m_{B}}.
\end{eqnarray}
where $n_{0}=|\psi_{0}|^{2}$ is the condensate density. 

As pointed out in \cite{Bruderer2008}, the coefficients $u_{\mathbf{q}}\left(\mathbf{r}\right)$,
$v_{\mathbf{q}}\left(\mathbf{r}\right)$ and Bogoliubov phonon spectrum
$\omega_{\mathbf{q}}$ are not changed by the presence of the impurity,
although the equilibrium positions of the modes are modified by $\vartheta\left(\mathbf{r}\right)$.
In \cite{Bruderer2008}, they express $\hat{\vartheta}\left(\mathbf{r}\right)=\vartheta\left(\mathbf{r}\right)+\hat{\zeta}\left(\mathbf{r}\right)$
in terms of Bogoliubov modes around the state $\psi_{0}\left(\mathbf{r}\right)$
as
\begin{equation}
\hat{\vartheta}\left(\mathbf{r}\right)=\sum_{\mathbf{q}}\left[u_{\mathbf{q}}e^{i\mathbf{q}\cdot\mathbf{r}}\hat{b}_{\mathbf{q}}-v_{\mathbf{q}}^{*}e^{-i\mathbf{q}\cdot\mathbf{r}}\hat{b}_{\mathbf{q}}^{\dagger}\right].
\end{equation}
In contrast to Eq. (\ref{eq:phonon_beta}), the bosonic operators
$\hat{b}_{\mathbf{q}}^{\dagger}$ ($\hat{b}_{\mathbf{q}}$) create
(annihilate) a quasi-particle around the ground state $\psi_{0}\left(\mathbf{r}\right)$
in absence of impurity rather than around state $\psi_{0}\left(\mathbf{r}\right)+\vartheta\left(\mathbf{r}\right)$.
The Hamiltonian (up to constant terms) for this effective phonon bath
can be simply expressed as 
\begin{equation}
\hat{H}_{B}=\sum_{\mathbf{q}}\hbar\omega_{\mathbf{q}}\hat{b}_{\mathbf{q}}^{\dagger}\hat{b}_{\mathbf{q}},
\end{equation}
if the impurity-BEC coupling $g_{IB}$ satisfies condition $|\langle\vartheta\left(\mathbf{r}\right)\rangle|\ll\psi_{0}\left(\mathbf{r}\right)$,
which implies the dimensionless relation: 
\begin{equation}
\left(|g_{IB}|/g_{B}\right)\ll n_{0}\xi^{D}\label{eq:couling_g_C}
\end{equation}
where $\xi\equiv\hbar/\sqrt{m_{B}g_{B}n_{0}}$ is the condensate healing
length and $D=3$ for a 3D homogeneous bosonic bath \cite{Bruderer2007}.
Strictly speaking, even when the coupling $g_{IB}$ is stronger than
this limit, the perturbation theory still qualitatively applies. For
very strong interactions, however, the bath can not be properly described
by Bogoliubov quasi-particles any more. The above perturbative method
can also be used in a system where the impurities and BEC are both
trapped by optical lattices \cite{Privitera2010}.

\subsection{Atom-phonon coupling}

The impurity-BEC interaction of the Hamiltonian can be written as
\begin{eqnarray}
\hat{H}_{\text{int}} & = & \sum_{\mathbf{q}}\sum_{\alpha,\beta}\sum_{i,j}\hbar\omega_{\mathbf{q}}M_{i,j;\mathbf{q}}^{\alpha\beta}\hat{b}_{\mathbf{q}}\hat{a}_{i}^{\alpha\dagger}\hat{a}_{j}^{\beta}+h.c.\label{eq:interaction-1}\\
M_{i,j;\mathbf{q}}^{\alpha\beta} & = & \frac{g_{IB}}{\hbar\omega_{\mathbf{q}}}\sqrt{\frac{n_{0}}{\Omega}}\left(u_{\mathbf{q}}-v_{\mathbf{q}}\right)m_{i,j;\mathbf{q}}^{\alpha\beta},
\end{eqnarray}
with $\alpha,\beta$ indicating the impurity Bloch bands and $m_{i,j;\mathbf{q}}^{\alpha\beta}\equiv\int d^{3}\mathbf{r}e^{i\mathbf{q}\cdot\mathbf{r}}W_{i}^{\alpha*}\left(\mathbf{r}\right)W_{j}^{\beta}\left(\mathbf{r}\right)$.
This term describes the impurity coupling to the phonon bath by creating
or annihilating phonons. The non-local coupling terms with $i\neq j$
are highly suppressed due to the local form of the interaction which
requires overlap between localized Wannier functions. For these reasons,
the above integrals $m_{i,j;\mathbf{q}}^{\alpha\beta}$ can be well
approximated by $\delta_{ij}m_{\mathbf{q}}^{\alpha\beta}e^{i\mathbf{q}\cdot\mathbf{R}_{i}}$
such that the value depends only on $\alpha,\beta$ and $\mathbf{q}$.
The dimensionless impurity-phonon coupling $M_{i;\mathbf{q}}^{\alpha\beta}$
can also be written as $M_{\mathbf{q}}^{\alpha\beta}e^{i\mathbf{q}\cdot\mathbf{R}_{i}}$
with 
\begin{equation}
M_{\mathbf{q}}^{\alpha\beta}\equiv g_{IB}\sqrt{\frac{n_{0}\epsilon_{\mathbf{q}}}{\Omega\left(\hbar\omega_{\mathbf{q}}\right)^{3}}}m_{\mathbf{q}}^{\alpha\beta}.
\end{equation}
Note that the impurity-phonon coupling $M_{i,\mathbf{q}}^{\alpha\beta}$
obeys:
\begin{equation}
\left(M_{i,\mathbf{q}}^{\alpha\beta}\right)^{*}=M_{i,-\mathbf{q}}^{\alpha\beta}.\label{eq:M_relation}
\end{equation}
The integral factors $m_{\mathbf{q}}^{\alpha\beta}$ take the form
\[
m_{\mathbf{q}}^{\alpha\beta}=\int d^{3}\mathbf{r}e^{i\mathbf{q}\cdot\mathbf{r}}\varphi_{x}^{\alpha*}\left(x\right)\varphi_{x}^{\beta}\left(x\right)|\phi^{0}(y)|^{2}|\phi^{0}(z)|^{2}.
\]
In a deep optical lattice, where the impurity Wannier functions in
Eqs. (\ref{eq:wannier_y}-\ref{eq:wannier_x1}) can be approximated
as harmonic oscillator states, we can explicitly evaluate these factors:
\begin{alignat}{1}
m_{\mathbf{q}}^{00}= & e^{-\left(\sigma_{\perp}^{2}q_{\perp}^{2}+\sigma_{x}^{2}q_{x}^{2}\right)/4},\nonumber \\
m_{\mathbf{q}}^{01}= & e^{-\left(\sigma_{\perp}^{2}q_{\perp}^{2}+\sigma_{x}^{2}q_{x}^{2}\right)/4}\left(iq_{x}\sigma_{x}/\sqrt{2}\right)=m_{\mathbf{q}}^{10},\nonumber \\
m_{\mathbf{q}}^{11}= & e^{-\left(\sigma_{\perp}^{2}q_{\perp}^{2}+\sigma_{x}^{2}q_{x}^{2}\right)/4}\left(1-q_{x}^{2}\sigma_{x}^{2}/2\right),\label{eq:coupling_factor}
\end{alignat}
by using the identity for Hermite polynomial integrals \cite{Daley2004}.
Here $q_{\perp}\equiv\sqrt{q_{y}^{2}+q_{z}^{2}}$ indicates the transverse
phonon momentum. 

In order to describe polaronic effects resulting from the impurity-BEC
coupling, we introduce a dimensionless coupling constant as in \cite{Grusdt2014}
and name it $\kappa$. This constant depends on the impurity-boson
and boson-boson interactions $g_{IB}$, $g_{BB}$ and the condensate
parameters $\xi$ and $m_{B}$ as
\begin{equation}
\kappa\equiv\sqrt{\frac{g_{IB}^{2}m_{B}}{g_{B}\hbar^{2}\xi}}.\label{eq:couling_kappa}
\end{equation}
As described in \cite{Grusdt2014}, this constant is the ratio $\kappa=E_{IB}/E_{{\rm ph}}$
between the characteristic impurity-boson interaction $E_{IB}=g_{IB}\sqrt{n_{0}\xi^{-3}}$
and the typical phonon energy $E_{{\rm ph}}=\hbar c/\xi$, where $c=\sqrt{g_{B}n_{0}/m_{B}}$
is the condensate speed of sound. In \cite{Tempere2009} an alternative
dimensionless coupling constant is used as
\begin{equation}
\alpha\equiv\frac{a_{IB}^{2}}{a_{BB}\xi},
\end{equation}
or equivalently $\alpha=4\pi n_{0}a_{IB}^{2}\xi$. These two coupling
constants are related by $\alpha=\left(\kappa\mu/m_{B}\right)^{2}/\pi$,
where $\mu$ is the reduced mass. There are also other coupling constants
used \cite{Cucchietti2006,Grusdt2014a}, which are slightly different
from $\alpha$ or $\kappa$. By tuning the impurity-boson scattering
length $a_{IB}^{\text{eff}}$, the coupling constant $\kappa$ can
be tuned continuously. However, the condition in eq.~(\ref{eq:couling_g_C})
requires that the coupling constant satisfies the relation 
\begin{equation}
\kappa\ll\kappa_{c}\equiv\frac{1}{2\sqrt{\pi}}\sqrt{\frac{\xi}{a_{B}}}=\left(64\pi^{3}a_{B}^{3}n_{0}\right)^{-\frac{1}{4}}.\label{eq:kappa_C}
\end{equation}
This upper limit of the coupling constant $\kappa_{c}$ depends only
on the boson-boson scattering length $a_{B}$ and the condensate density
$n_{0}$ instead of the specific mass ratio $m_{I}/m_{B}$. In a typical
BEC system, this maximum coupling constant is relatively small, such
as $\kappa_{c}\approx2.4$ for $^{87}\text{Rb}$ with $a_{B}=100a_{0}$
and $n_{0}\approx10^{14}\text{cm}^{-3}$. In order to reach larger
values of $\kappa_{c}$ in realistic systems, one need to reduce the
condensate density $n_{0}$ or the Bose-Bose scattering length $a_{B}$. 

Finally the effective Hamiltonian with two-band Fr\"{o}hlich impurity-phonon
coupling is 
\begin{alignat}{1}
\hat{H}= & -\sum_{\langle i,j\rangle}\sum_{\alpha}J^{\alpha}\hat{a}_{i}^{\alpha\dagger}\hat{a}_{j}^{\alpha}+\sum_{i}\sum_{\alpha}\varepsilon^{\alpha}\hat{n}_{i}^{\alpha}\nonumber \\
+ & \sum_{\mathbf{q}}\hbar\omega_{\mathbf{q}}\hat{b}_{\mathbf{q}}^{\dagger}\hat{b}_{\mathbf{q}}\nonumber \\
+ & \sum_{i,\mathbf{q}}\sum_{\alpha,\beta}\hbar\omega_{\mathbf{q}}M_{i,\mathbf{q}}^{\alpha\beta}\left(\hat{b}_{\mathbf{q}}+\hat{b}_{-\mathbf{q}}^{\dagger}\right)\hat{a}_{i}^{\alpha\dagger}\hat{a}_{i}^{\beta}.\label{eq:Hamil_0}
\end{alignat}
Here we used the relations in Eq.~(\ref{eq:M_relation}). This Hamiltonian
describes a general two-band system with impurity-phonon coupling,
which can also be realized by other experimental setups such as hybrid
atom-ion systems \cite{Bissbort2013}.

\section{Variational two-band Polaron transformation\label{sec:Variational-Polaron-transformati}}

\subsection{Transformation with exponential quadratic operators}

This two-band Hamiltonian in Eq.~(\ref{eq:Hamil_0}), with Fr\"{o}hlich-type
impurity-phonon coupling, can not be solved analytically even for
the case of a single impurity. The goal of this paper is to find a
simple but non-trivial variational method which can deal with the
two-band system (\ref{eq:Hamil_0}) in general. We choose the Lang-Firsov
polaron transformation approach and generalize it to two-band system.
This canonical transformation is exact and decouples the impurity-BEC
interaction term in a new quasi-particle basis. In this basis, the
kinetic part in Hamiltonian (\ref{eq:Hamil_0}) contains the dynamics
of quasiparticle and its interactions between the transformed phonon
bath. We solve the coherent part of this transformed Hamiltonian with
a variational treatment and take into account the remaining incoherent
parts by a master equation. 

Firstly we introduce the basic concept of the transformation for a
single band before we extend it to the two-band case. When a single
impurity moves in a lattice and couples to a phonon bath, there are
exact solutions in both the weak and strong coupling limit \cite{Mahan2000}.
When the impurity-phonon interaction is much weaker than the impurity
kinetic energy, the impurity behaves as a free particle in a lattice.
On the other hand, when the interaction is much larger than the kinetic
part, the impurity will be tightly dressed by a ``cloud'' of phonons,
forming a quasi-particle. The phonons are tied to the impurity such
that the impurity cannot move on its own but must drag around a phonon
cloud. This increases the effective mass of the quasiparticle. In
the intermediate coupling region, the phonon dressing competes with
the impurity dynamics. In order to describe this competition, a variational
ground state can be used to connect between the weak and strong coupling
limits \cite{Harris1985,Chatterjee2000,Stojanovic2004,Barone2006,Agarwal2013,Benjamin2014}.
This variational ansatz is equivalent to a canonical transformation
$\tilde{H}\equiv e^{\hat{S}}\hat{H}e^{-\hat{S}}$ with $\hat{S}\equiv\sum_{i,\mathbf{q}}\Lambda_{i,\mathbf{q}}\left(\hat{b}_{-\mathbf{q}}^{\dagger}-\hat{b}_{\mathbf{q}}\right)\hat{n}_{i}$
where $\Lambda_{i,\mathbf{q}}$ are the variational parameters. The
transformed Hamiltonian $\tilde{H}$ still cannot be solved analytically,
but can be separated into a coherent part $\langle\tilde{H}\rangle_{T}$
and an incoherent part $\tilde{H}_{\text{inc}}\equiv\tilde{H}-\langle\tilde{H}\rangle_{T}$
where $\langle\cdots\rangle_{T}$ indicates a thermal average over
the phonon bath. The coherent part, which is decoupled from the phonon
bath, is of the form of an extended (polaronic) Hubbard model. The
incoherent part describes the residual coupling between polaron quasi-particle
and phonon bath. Compared to the initial ``bare'' impurity-phonon
coupling, this incoherent part is significantly reduced by the polaron
transformation. We first focus on the coherent part and neglect the
incoherent terms. The variational parameters $\Lambda_{i,\mathbf{q}}$
are determined by minimizing the coherent Hamiltonian energy and approach
$\Lambda_{i,\mathbf{q}}=M_{i,\mathbf{q}}$ in the strong coupling
limit. Finally, the residual incoherent part can be included by a
perturbative approach such as the Lindblad master equation. 

In order to find a suitable variational transformation for the two-band
system, we modify the Lang-Firsov polaron transformation $\tilde{H}=e^{\hat{S}}He^{-\hat{S}}$
by extending the impurity-phonon interaction to the two-band form.
It takes the form: 
\begin{equation}
\hat{S}\equiv\left[\sum_{i,\mathbf{q}}\sum_{\alpha,\beta}\Lambda_{i,\mathbf{q}}^{\alpha\beta}\left(\hat{b}_{-\mathbf{q}}^{\dagger}-\hat{b}_{\mathbf{q}}\right)\hat{a}_{i}^{\alpha\dagger}\hat{a}_{i}^{\beta}\right],\label{eq:transformation_S}
\end{equation}
with $\Lambda_{i,\mathbf{q}}^{\alpha\beta}\equiv\Lambda_{\mathbf{q}}^{\alpha\beta}e^{i\mathbf{q}\cdot\mathbf{R}_{i}}$.
A similar method was also applied by Sibley and Munn in \cite{Silbey1980,Munn1985a,Munn1985b,Chen2011}
and Stojanovi$\acute{\textrm{c}}$ \textit{et al.} in \cite{Stojanovic2004}
for a single-band system with non-local impurity-phonon coupling.
Our initial guess for the variational parameters is the coupling element
itself, $M_{\mathbf{q}}^{\alpha\beta}$, and we constrain the variational
parameters to obey the same symmetry properties as $M_{\mathbf{q}}^{\alpha\beta}$,
namely that of Eq.~(\ref{eq:M_relation}). By using the Baker-Campbell-Hausdorff
formula, $e^{\hat{S}}\hat{A}e^{-\hat{S}}=\hat{A}+\left[\hat{S},\;\hat{A}\right]+\frac{1}{2!}\left[\hat{S},\left[\hat{S},\;\hat{A}\right]\right]+\cdots$
, we can\textcolor{red}{{} }derive the transformed Hamiltonian with
exponential quadratic operators as outlined in Appendix \ref{sub:App_Polaron-transformation}.
For convenience, these expressions can be written in $2\times2$ matrix
form via
\[
\mathbf{\Lambda}_{\mathbf{q}}\equiv\left(\begin{array}{cc}
\Lambda_{\mathbf{q}}^{00} & \Lambda_{\mathbf{q}}^{01}\\
\Lambda_{\mathbf{q}}^{10} & \Lambda_{\mathbf{q}}^{11}
\end{array}\right);\;\hat{\mathbf{b}}_{\mathbf{\mathbf{q}}}\equiv\left(\begin{array}{cc}
\hat{b}_{\mathbf{q}} & 0\\
0 & \hat{b}_{\mathbf{q}}
\end{array}\right),
\]
and 
\[
\mathbf{M}_{i,\mathbf{q}}\equiv\left(\begin{array}{cc}
M_{\mathbf{\mathbf{q}}}^{00} & M_{\mathbf{\mathbf{q}}}^{01}\\
M_{\mathbf{\mathbf{q}}}^{10} & M_{\mathbf{\mathbf{q}}}^{11}
\end{array}\right)e^{i\mathbf{q}\cdot\mathbf{R}_{i}}.
\]

After the transformation, the impurity annihilation and creation operators
can be expressed through
\begin{equation}
e^{\hat{S}}\hat{a}_{i}^{\alpha}e^{-\hat{S}}=\sum_{\beta}\left(\hat{\mathbf{X}}_{i}\right)_{\alpha\beta}\hat{a}_{i}^{\beta},\label{eq:transform_a}
\end{equation}
and similarly for the creation operators. The matrix operators $\hat{\mathbf{X}}_{i}$
are found to be
\begin{equation}
\hat{\mathbf{X}}_{i}\equiv e^{-\sum_{\mathbf{q}}\mathbf{\mathbf{\Lambda}}_{\mathbf{\mathbf{q}}}e^{i\mathbf{\mathbf{q}}\cdot\mathbf{R}_{i}}\left(\hat{\mathbf{b}}_{-\mathbf{q}}^{\dagger}-\hat{\mathbf{b}}_{\mathbf{\mathbf{q}}}\right)}.\label{eq:X_operators}
\end{equation}
In the same fashion as was described above for the single band system,
this canonical transformation is equivalent to defining a new quasi-particle,
which represents an impurity dressed by the phonon cloud forming a
polaron. This transformation also shifts the equilibrium position
of the phonon bath by 
\begin{equation}
e^{\hat{S}}\hat{b}_{\mathbf{\mathbf{q}}}e^{-\hat{S}}=\hat{b}_{\mathbf{q}}+\sum_{i,\alpha,\beta}\left(\hat{\mathbf{X}}_{i}^{\dagger}\hat{\mathbf{b}}_{\mathbf{\mathbf{q}}}\hat{\mathbf{X}}_{i}-\hat{\mathbf{b}}_{\mathbf{\mathbf{q}}}\right)_{\alpha\beta}\hat{a}_{i}^{\alpha\dagger}\hat{a}_{i}^{\beta},\label{eq:transform_b}
\end{equation}
but it does not modify the phonon dispersion relation. In Appendix
\ref{sub:App_Polaron-transformation}, we show the derivations for
Eq.~(\ref{eq:transform_a}-\ref{eq:transform_b}). The transformed
polaronic Hamiltonian can be written as:\begin{widetext} 
\begin{alignat}{1}
\tilde{H}= & -\sum_{\langle i,j\rangle}\sum_{\alpha\beta}\left(\hat{\mathbf{X}}_{i}^{\dagger}\mathbf{J}\hat{\mathbf{X}}_{j}\right)_{\alpha\beta}\hat{a}_{i}^{\alpha\dagger}\hat{a}_{j}^{\beta}+\sum_{i}\sum_{\alpha\beta}\left(\hat{\mathbf{X}}_{i}^{\dagger}\mathbf{\varepsilon}\hat{\mathbf{X}}_{i}\right)_{\alpha\beta}\hat{a}_{i}^{\alpha\dagger}\hat{a}_{i}^{\beta}+\sum_{\mathbf{q}}\hbar\omega_{\mathbf{\mathbf{q}}}\hat{b}_{\mathbf{\mathbf{q}}}^{\dagger}\hat{b}_{\mathbf{\mathbf{q}}}\nonumber \\
+ & \sum_{i,\alpha,\beta}\sum_{\mathbf{\mathbf{q}}}\hbar\omega_{\mathbf{\mathbf{q}}}\left[\left(\hat{\mathbf{X}}_{i}^{\dagger}\hat{\mathbf{b}}_{\mathbf{\mathbf{q}}}^{\dagger}\mathbf{M}_{i,\mathbf{\mathbf{q}}}^{\dagger}\hat{\mathbf{X}}_{i}\right)+\left(\hat{\mathbf{X}}_{i}^{\dagger}\mathbf{M}_{i,\mathbf{\mathbf{q}}}\hat{\mathbf{b}}_{\mathbf{\mathbf{q}}}\hat{\mathbf{X}}_{i}\right)+\left(\hat{\mathbf{X}}_{i}^{\dagger}\hat{\mathbf{b}}_{\mathbf{\mathbf{q}}}^{\dagger}\hat{\mathbf{b}}_{\mathbf{q}}\hat{\mathbf{X}}_{i}\right)-\hat{\mathbf{b}}_{\mathbf{q}}^{\dagger}\hat{\mathbf{b}}_{\mathbf{\mathbf{q}}}\right]_{\alpha\beta}\hat{a}_{i}^{\alpha\dagger}\hat{a}_{i}^{\beta}\nonumber \\
+ & \sum_{i,\alpha,\beta}\sum_{j,\alpha',\beta'}\sum_{\mathbf{q}}\frac{\hbar\omega_{\mathbf{q}}}{2}\left[\left(\hat{\mathbf{X}}_{i}^{\dagger}\hat{\mathbf{b}}_{\mathbf{q}}^{\dagger}\hat{\mathbf{X}}_{i}-\hat{\mathbf{b}}_{\mathbf{\mathbf{q}}}^{\dagger}\right)_{\alpha\beta}\left(2\hat{\mathbf{X}}_{j}^{\dagger}\mathbf{M}_{j,\mathbf{q}}^{\dagger}\hat{\mathbf{X}}_{j}+\hat{\mathbf{X}}_{j}^{\dagger}\hat{\mathbf{b}}_{\mathbf{q}}\hat{\mathbf{X}}_{j}-\hat{\mathbf{b}}_{\mathbf{q}}\right)_{\alpha'\beta'}+h.c.\right]\hat{a}_{i}^{\alpha\dagger}\hat{a}_{j}^{\alpha'\dagger}\hat{a}_{j}^{\beta'}\hat{a}_{i}^{\beta},\label{eq:transformed-Hamil_0}
\end{alignat}
\end{widetext}where $\mathbf{J}$ and $\mathbf{\varepsilon}$ is
the (diagonal) matrix form of $J^{\alpha}$ and $\varepsilon^{\alpha}$.
Here we combine all single impurity contributions in the second line.
The last term, which is zero when only a single impurity is considered,
describes induced polaron-polaron interactions due to the coupling
with the phonon bath. Compared to the original Hamiltonian in Eq.~(\ref{eq:Hamil_0}),
the hopping and on-site energy have been modified. As for the single
band system, this Hamiltonian contains all interactions exactly and
is hard to solve analytically. Motivated by the single-band system,
we separate the Hamiltonian into coherent and incoherent parts, $\langle\tilde{H}\rangle_{T}$
and $\tilde{H}-\langle\tilde{H}\rangle_{T}$ respectively. The explicit
form of the coherent terms, which conserve the number of phonons,
is determined in Appendix \ref{sub:App_Polaron-transformation}. In
Eqs.\ (\ref{eq:thermal_1_end}, \ref{eq:thermal_2}, \ref{eq:thermal_3},
\ref{eq:thermal_4}) we calculate all possible coherent terms in Eq.~(\ref{eq:transformed-Hamil_0}).
After determining the variational parameters $\Lambda_{\mathbf{q}}$
by minimizing the free energy of the coherent part, we treat the residual
polaron-bath coupling in the incoherent part as a perturbation and
solve it by a Lindblad master equation.

\subsection{Diagonal transformation matrix}

In Appendix \ref{sub:App_Polaron-transformation}, we calculate the
coherent part $\langle\tilde{H}\rangle_{T}$ by averaging over the
phonon bath and assuming it is thermal. In contrast to the single-band
case, these calculations are demanding when both intra- and inter-band
phonon couplings are included. In order to determine the variational
parameters $\mathbf{\Lambda}_{\mathbf{q}}$, we finally need to minimize
the free energy of whole coherent Hamiltonian. Until now, we did not
make any assumptions for our variational parameters $\mathbf{\Lambda}_{\mathbf{q}}$
except for the symmetry relations in Eq.~(\ref{eq:M_relation}). 

Unfortunately, the general result of the transformed Hamiltonian in
Eq.~(\ref{eq:transformed-Hamil_0}) and its corresponding coherent
part are still quite complicated. It can be further simplified by
making some approximations suitable to our specific system. Due to
conservation of energy, the phonon-induced inter-band dynamics requires
the phonon energy to match the impurity band gap, i.e. $\hbar\omega_{\mathbf{q}}\approx\varepsilon^{\Delta}$.
This energy scale involves a phonon with particle-like dispersion
and momentum $|\mathbf{q}|\thickapprox\sqrt{2m_{B}\varepsilon^{\Delta}}/\hbar$
significantly far from zero. The inter-band coupling $M_{\mathbf{q}}^{01}$
for this large phonon momentum is highly reduced due to the Gaussian
decay of $m_{\mathbf{q}}^{01}$ in Eq.~(\ref{eq:coupling_factor}).
On the other hand, intra-band dynamics requires a phonon energy $\hbar\omega_{\mathbf{q}}\approx J^{0};J^{1}$
with phonon-like dispersion and small momentum $|\mathbf{q}|\thickapprox J^{\alpha}\sqrt{m_{B}/\left(g_{B}n_{0}\right)}/\hbar=J^{\alpha}/\left(\hbar c\right)$.
The intra-band coupling $M_{\mathbf{q}}^{\alpha\alpha}$ is not reduced
too much at this smaller momentum. In the polaron transformation,
the parameters $\Lambda_{\mathbf{\mathbf{q}}}^{\alpha\beta}$ reflect
the dressing of the impurity by phonons and are closely related to
$M_{\mathbf{q}}^{\alpha\beta}$. For this reason, we treat the inter-band
coupling as small and approximate the matrices $\mathbf{\Lambda}_{\mathbf{q}}$
as diagonal: 
\[
\mathbf{\mathbf{\Lambda}}_{\mathbf{q}}=\left[\begin{array}{cc}
\lambda_{\mathbf{\mathbf{q}}}^{0}M_{\mathbf{\mathbf{q}}}^{00} & 0\\
0 & \lambda_{\mathbf{q}}^{1}M_{\mathbf{q}}^{11}
\end{array}\right],
\]
with variational parameters $\lambda_{\mathbf{q}}^{0}$ and $\lambda_{\mathbf{q}}^{1}$.
Since the intra-band couplings $M_{\mathbf{q}}^{\alpha\alpha}$ are
purely real numbers in our system, we assume $\lambda_{\mathbf{q}}^{\alpha}$
are also real numbers. After the transformation with these diagonal
matrices $\mathbf{\mathbf{\mathbf{\Lambda}}}_{\mathbf{q}}$, we have
then decoupled the intra-band impurity-phonon coupling and leave the
(relatively) small inter-band coupling in the new polaronic two-band
Hamiltonian. The coherent part, with phonons eliminated by thermal
averaging, is a many-body Hamiltonian:
\begin{alignat}{1}
\langle\tilde{H}\rangle_{T}= & -\sum_{\langle i,j\rangle,\alpha}J_{{\rm P}}^{\alpha}\hat{a}_{i}^{\alpha\dagger}\hat{a}_{j}^{\alpha}+\sum_{i,\alpha}\varepsilon_{P}^{\alpha}\hat{n}_{i}^{\alpha}+\sum_{\mathbf{q}}\hbar\omega_{\mathbf{q}}\langle\hat{b}_{\mathbf{q}}^{\dagger}\hat{b}_{\mathbf{q}}\rangle_{T}\nonumber \\
 & +\hat{V}_{{\rm P}},\label{eq:coherent}
\end{alignat}
with renormalized polaronic hopping terms $J_{{\rm P}}^{\alpha}$
and on-site energy $\varepsilon_{{\rm P}}^{\alpha}$ including the
polaron energy shift:
\begin{alignat}{1}
J_{{\rm P}}^{\alpha}\equiv & J^{\alpha}\langle\left(\hat{\mathbf{X}}_{i}\right)_{\alpha\alpha}^{\dagger}\left(\hat{\mathbf{X}}_{j}\right)_{\alpha\alpha}\rangle_{T},\nonumber \\
\varepsilon_{{\rm P}}^{\alpha}\equiv & \varepsilon^{\alpha}-\sum_{\mathbf{q}}\hbar\omega_{\mathbf{q}}\left(\Lambda_{\mathbf{\mathbf{q}}}^{\alpha\alpha}\right)\left(2M_{\mathbf{\mathbf{q}}}^{\alpha\alpha}-\Lambda_{\mathbf{q}}^{\alpha\alpha}\right)^{*}.\label{eq:J_eff}
\end{alignat}
In the above formula, we use the fact that operators $\hat{\mathbf{X}}_{i}$
have only diagonal terms. There are also induced interactions $\hat{V}_{{\rm P}}$
between multiple polarons, which can lead to strong correlations in
the system and will be discussed in the next section. In the above
calculations, we need to sum over all possible phonon momenta $\mathbf{q}$.
In the thermodynamic limit of the phonon bath, we use the relation
$\sum_{\mathbf{q}}\rightarrow\frac{\Omega}{\left(2\pi\right)^{D}}\int d\mathbf{q}$
with quantization volume $\Omega$ for the phonons, and write this
explicitly in cylindrical coordinates: $\frac{\Omega}{\left(2\pi\right)^{3}}\int dq_{\perp}\int dq_{x}2\pi q_{\perp}.$ 

For the polaronic intra-band hopping $J_{{\rm P}}^{\alpha}$, we only
need to calculate nearest-neighbor terms with $j=i\pm1$ regardless
of the specific value of $i$. By noting that 
\begin{equation}
\langle\left(\hat{\mathbf{X}}_{i}\right)_{\alpha\alpha}^{\dagger}\left(\hat{\mathbf{X}}_{j}\right)_{\beta\beta}\rangle_{T}=e^{-\sum_{\mathbf{q}}\left(N_{\mathbf{q}}+1/2\right)|\Lambda_{i,\mathbf{q}}^{\alpha\alpha}-\Lambda_{j,\mathbf{q}}^{\beta\beta}|^{2}},
\end{equation}
where $N_{\mathbf{q}}\equiv\left(\exp\left(\hbar\omega_{\mathbf{q}}/k_{\text{B}}T\right)-1\right)^{-1}$
is the thermally averaged phonon occupation number, we define polaronic
renormalization factors 
\begin{equation}
S_{T}^{\alpha}\equiv\sum_{\mathbf{q}}\left(2N_{\mathbf{q}}+1\right)\left[1-\cos\left(q_{x}\cdot d\right)\right]|\Lambda_{\mathbf{q}}^{\alpha\alpha}|^{2}
\end{equation}
for each band and thus $J_{{\rm P}}^{\alpha}=J^{\alpha}\exp\left(-S_{T}^{\alpha}\right)$.\textcolor{red}{{} }

We would like to note that the choice of diagonal transformation matrices
$\mathbf{\mathbf{\mathbf{\Lambda}}}_{\mathbf{q}}$ works well when
the inter-band coupling is less important that the intra-band coupling.
By choosing diagonal matrices, we treat each of the two bands independently.
This approach is similar as the authors in Ref. \cite{Bruderer2007,Bruderer2008}
did for single band problem, except a variational treatment was used.
On the other hand, when the off-diagonal terms in the coupling matrix
$\mathbf{M}_{\mathbf{q}}$ are comparable or larger than the diagonal
terms, one must instead deal with the general transformed Hamiltonian
in Eq.~(\ref{eq:transformed-Hamil_0}). For completeness, we describe
a more general approximate technique in Appendix \ref{sub:App_Polaron-transformation},
which may be useful for other types of systems. We intend to address
the approximation introduced in this section more explicitly and quantitatively
in a future publication.

\section{Coherent polaron dynamics and interactions\label{sec:coherent-part}}

\subsection{Single polaron band structures}

In the previous section, we derived a general form of the two-band
polaron transformation and calculated the resulting coherent part
of the Hamiltonian in Eq.~(\ref{eq:coherent}). For the system with
a single polaron, there are only intra-band terms in the coherent
Hamiltonian (\ref{eq:coherent}). The single impurity is dressed by
a coherent phonon cloud in each band. The residual inter-band polaron-bath
coupling will appear only in the incoherent Hamiltonian. It is easy
to diagonalize the single-polaron coherent part in the momentum representation
and minimize the free energy $F\equiv-k_{\text{B}}T\ln\sum_{\alpha,k}\exp\left(-E_{k}^{\alpha}/k_{\text{B}}T\right)$
for this two-band system, with polaron dispersion: 
\begin{equation}
E_{k}^{\alpha}\equiv2J_{{\rm P}}^{\alpha}\cos\left(k\cdot d\right)+\varepsilon_{{\rm P}}^{\alpha}.
\end{equation}
Here $k$ is quasi-momentum in longitudinal direction. These variational
parameters, which are real numbers, can then be determined by the
self-consistent equations: 
\begin{equation}
\lambda_{\mathbf{q}}^{\alpha}=\frac{\sum_{k,\alpha'}\exp\left(-E_{k}^{\alpha'}/k_{\text{B}}T\right)}{\sum_{k,\alpha'}\left[1-2J_{{\rm P}}^{\alpha}f_{\mathbf{q}}\cos(k\cdot d)/\omega_{\mathbf{q}}\right]\exp\left(-E_{k}^{\alpha'}/k_{\text{B}}T\right)}\label{eq:lambda}
\end{equation}
with $f_{\mathbf{q}}\equiv\left(2N_{\mathbf{q}}+1\right)\left[1-\cos\left(q_{x}d\right)\right]$. 

Before numerically calculating the variational parameters $\lambda_{\mathbf{q}}^{\alpha}$,
we firstly discuss some properties of this self-consistent equation.
Considering a simplified model with momentum-independent variational
parameters $\lambda$ in a single band system, we choose to minimize
only the ground state energy. The self-consistent equation (\ref{eq:lambda})
will be modified as
\begin{equation}
\lambda=\left[1+2|J|\frac{\sum_{\mathbf{q}}f_{\mathbf{q}}|M_{\mathbf{q}}|^{2}}{\sum_{\mathbf{q}}\hbar\omega_{\mathbf{q}}|M_{\mathbf{q}}|^{2}}e^{-\lambda^{2}\sum_{\mathbf{q}}f_{\mathbf{q}}|M_{\mathbf{q}}|^{2}}\right]^{-1}.
\end{equation}
From the analysis in \cite{Stojanovic2004,Barone2006}, this equation
has two locally stable solutions $\lambda_{-},\lambda_{+}$ once the
\textit{adiabatic }regime is achieved when 
\begin{equation}
2|J|\frac{\sum_{\mathbf{q}}f_{\mathbf{q}}|M_{\mathbf{q}}|^{2}}{\sum_{\mathbf{q}}\hbar\omega_{\mathbf{q}}|M_{\mathbf{q}}|^{2}}>\frac{e^{3/2}}{2}.\label{eq:adiabatic}
\end{equation}
These two solutions $\lambda_{-},\lambda_{+}$, corresponding to two
local minima of ground state energy, indicate the impurity is respectively
loosely or tightly dressed by phonons. At a critical impurity-phonon
coupling with $\sum_{\mathbf{q}}f_{\mathbf{q}}|M_{\mathbf{q}}|^{2}=27/8$,
when the two minima become equal, the lowest energy state solution
abruptly switches from $\lambda_{-}$ to $\lambda_{+}$, indicating
a first-order\textit{ polaronic transition}. On the other hand, the
solution of $\lambda$ is a smooth and continuous crossover when the
adiabatic condition is broken with small value of $|J|.$ This transition-crossover
behavior also appears later when solving the self-consistent equation
(\ref{eq:lambda}) numerically, although we have considered a much
more simplified model here. Strictly speaking, this sharp polaronic
transition in the adiabatic regime is due to the mean-field approximation
by thermal averaging of the phonon degrees of freedom. This drawback
could be improved if we were to treat the incoherent dynamics properly,
by taking into account fluctuations or using a master equation method. 

\begin{figure}
\includegraphics[bb=0bp 0bp 250bp 262bp,clip,width=0.5\textwidth]{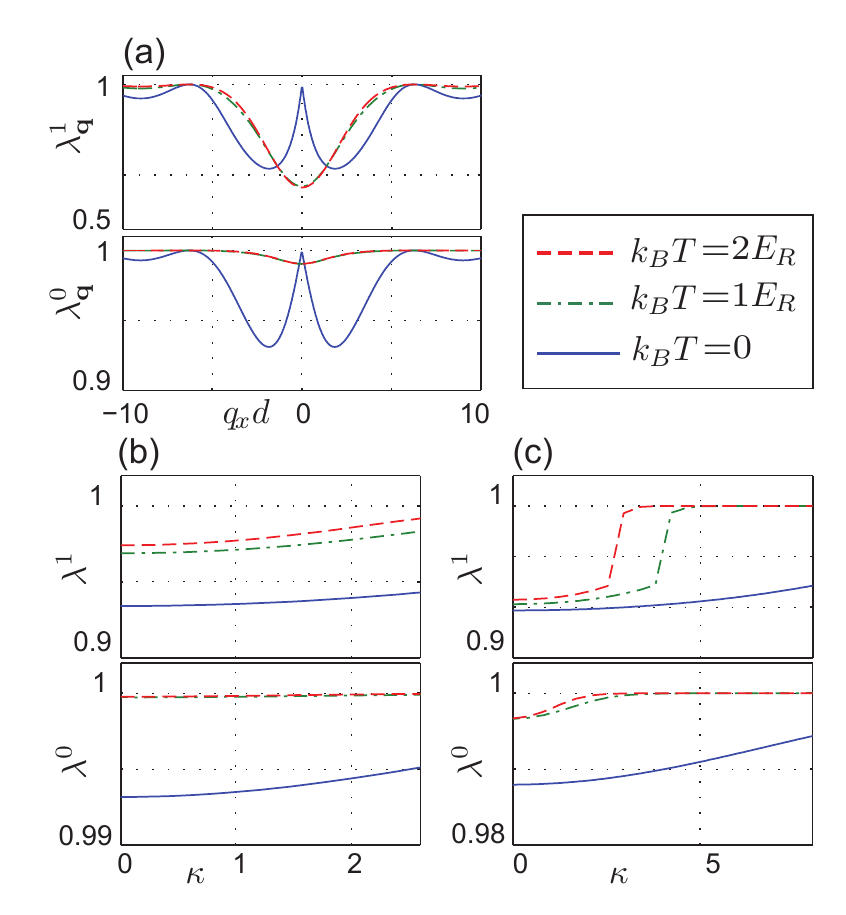}

\protect\caption{(a) Variational parameters $\lambda_{\mathbf{q}}^{\alpha}$ with $\mathbf{q}=\left(q_{x},0,0\right)$
and $\kappa=1$ in higher band (upper) and lower band (lower) at different
temperatures $k_{\text{B}}T=0,\,E_{R},\,2E_{R}$. (b)(c) Factors $\lambda^{\alpha}\equiv\sum_{\mathbf{q}}\Lambda_{\mathbf{\mathbf{q}}}^{\alpha\alpha}/\sum_{\mathbf{q}}M_{\mathbf{q}}^{\alpha\alpha}$
in each band at different temperatures. Maximum coupling constant
$\kappa_{c}=2.4$ for condensate density $n_{0}=1\times10^{14}\text{cm}^{-3}$
(b) and $\kappa_{c}=7.6$ for condensate density $n_{0}=0.01\times10^{14}\text{cm}^{-3}$(c).
Other parameters are: $m_{I}=133,\,m_{B}=87$; $V_{I}^{x}=9E_{R},\,V_{I}^{\perp}=25V_{I}^{x}$
and $a_{B}=100a_{0}$. \label{fig: variational lambda}}
\end{figure}

We now compare the variational parameters $\lambda_{\mathbf{q}}^{\alpha}=\Lambda_{\mathbf{q}}^{\alpha\alpha}/M_{\mathbf{q}}^{\alpha\alpha}$
from Eq.~(\ref{eq:lambda}) and polaron dressing effects for different
system parameters. In Fig.~\ref{fig: variational lambda}(a) we show
the coupling factor $\lambda_{\mathbf{q}}^{\alpha}$ in each band
for the temperatures $k_{\text{B}}T=0,\,E_{R},\,2E_{R}$, where it
can be seen that the variational parameters $\lambda_{\mathbf{q}}^{\alpha}$
are always smaller than 1. This shows the competition between intra-band
dynamics and phonon dressing effects. These variational parameters
are also different for the two bands, since the polaron is dressed
differently in each band. We have focused on the results for momentum
in the longitudinal direction with $\mathbf{q}=\left(q_{x},0,0\right)$
and find that the low-momentum phonons are less dressed at higher
temperatures $k_{\text{B}}T\approx E_{R}\gg J^{\alpha}$. Here $E_{R}$
corresponds to a temperature of about $65{\rm nK}$ for a $^{133}\text{Cs}$
impurity trapped by lasers with wavelength $1064{\rm nm}$. In Fig.~\ref{fig: variational lambda}(b,~c)
we also show the the factors $\lambda^{\alpha}\equiv\sum_{\mathbf{q}}\Lambda_{\mathbf{q}}^{\alpha\alpha}/\sum_{\mathbf{q}}M_{\mathbf{q}}^{\alpha\alpha}$.
These factors $\lambda^{\alpha}$ thus indicate the differences between
our initial guess for the variational parameters $\Lambda_{\mathbf{q}}^{\alpha\alpha}=M_{\mathbf{q}}^{\alpha\alpha}$,
as is sometimes used for the transformation, and the full minimization
of free energy for the variational parameters. As shown in Fig.~\ref{fig: variational lambda},
although these factors are different for each band and different temperatures,
they always approach $\lambda^{\alpha}=1$ in the limit of strong
interaction. In order to reach larger values of the coupling constant,
in Fig.~\ref{fig: variational lambda}(c) we show results with $\kappa_{c}=7.6$
by assuming condensate density $n_{0}=0.01\times10^{14}\text{cm}^{-3}$.
In the upper plot of Fig.~\ref{fig: variational lambda}(c), we notice
that the parameter $\lambda$ shows a polaronic transition in the
adiabatic regime when the condition Eq.~(\ref{eq:adiabatic}) is
satisfied at finite temperature. On the other hand, $\lambda$ shows
a smooth crossover behavior in the non-adiabatic regime at zero temperature.

\begin{figure*}
\includegraphics[bb=0bp 0bp 612bp 170bp,clip,width=1\textwidth]{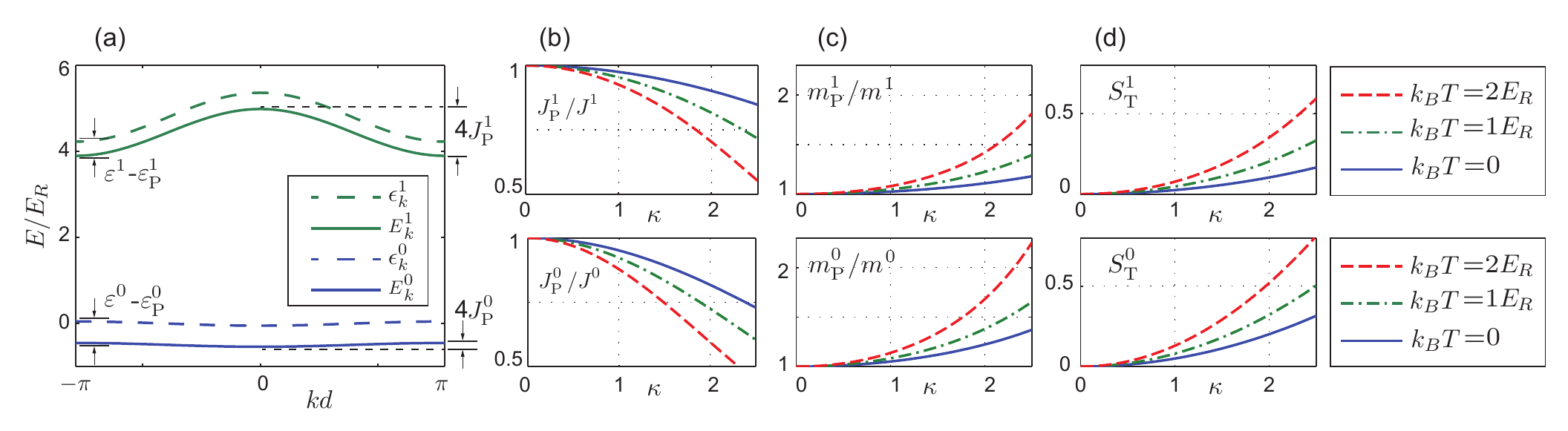}

\protect\caption{(a) Dispersion relation for bare impurity $\epsilon_{k}^{\alpha}$
and for polaron $E_{k}^{\alpha}$ with $\kappa=\kappa_{c}$; (b) renormalized
hopping $J_{{\rm P}}^{\alpha}/J^{\alpha}$; (c) polaron effective
mass $m_{{\rm P}}^{\alpha}/m_{0}^{\alpha}$; (d) renormalization factor
$S_{T}^{\alpha}$ for each band at different temperatures $k_{\text{B}}T=0;\,E_{R};\,2E_{R}$.
Other parameters are $m_{I}=133,\,m_{B}=87$; $V_{I}^{x}=9E_{R},\,V_{I}^{\perp}=25V_{I}^{x}$;
$a_{B}=100a_{0},\,n_{0}=1\times10^{14}\text{cm}^{-3}$ and $\kappa_{c}=2.4$.
\label{fig:7 E_k}}
\end{figure*}

In the coherent part of the Hamiltonian Eq.~(\ref{eq:coherent}),
the single polaron band structure is modified by phonon dressing effects
in Eq.~(\ref{eq:J_eff}). Effectively, the polaron is trapped in
a deeper lattice, with larger mass. The effective mass of a single
polaron (at $k_{0}=0$) in an optical lattice can be defined as
\begin{equation}
m_{{\rm P}}^{\alpha}\left(k_{0}\right)\equiv\hbar^{2}\left(\frac{\partial^{2}E_{k}^{\alpha}}{\partial k^{2}}\mid_{k_{0}}\right)^{-1}=\frac{\hbar^{2}}{2J_{{\rm P}}^{\alpha}}.
\end{equation}
If the impurity-BEC coupling $g_{IB}$ increases, the polaron effective
mass will increase exponentially as $m_{{\rm P}}^{\alpha}=m_{0}^{\alpha}\exp\left(S_{T}^{\alpha}\right)$,
with $m_{0}^{\alpha}$ indicating the impurity effective mass at $\kappa=0$.
In Fig.~\ref{fig:7 E_k} we compare energy spectrum $E_{k}^{\alpha}$,
renormalization of intra-band hopping $J_{{\rm P}}^{\alpha}/J^{\alpha}$,
polaron effective mass $m_{{\rm P}}^{\alpha}/m_{0}^{\alpha}$ and
renormalization factor $S_{T}^{\alpha}$ for each band at different
temperatures. 

\begin{figure}
\includegraphics[bb=0bp 0bp 280bp 132bp,clip,width=0.5\textwidth]{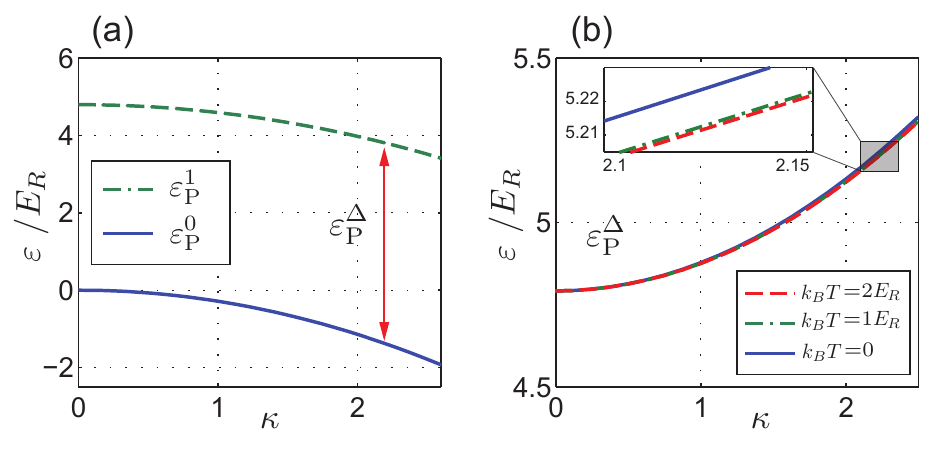}

\protect\caption{(a) On-site energy $\varepsilon_{{\rm P}}^{\alpha}$ including polaron
energy shift at zero temperature; (b) renormalized band gap $\varepsilon_{{\rm P}}^{\Delta}\equiv\varepsilon_{{\rm P}}^{1}-\varepsilon_{{\rm P}}^{0}$
at different temperatures $k_{\text{B}}T=0;\,E_{R};\,2E_{R}$. The
energy renormalization is almost temperature-independent. Other parameters
are $m_{I}=133,\,m_{B}=87$; $V_{I}^{x}=9E_{R},\,V_{I}^{\perp}=25V_{I}^{x}$;
$a_{B}=100a_{0},\,n_{0}=1\times10^{14}\text{cm}^{-3}$ and $\kappa_{c}=2.4$.
\label{fig:8 E_p-1}}
\end{figure}

Due to phonon dressing effects, the polaronic band gap $\varepsilon_{{\rm P}}^{\Delta}\equiv\varepsilon_{{\rm P}}^{1}-\varepsilon_{{\rm P}}^{0}$
is also increased. This will affect the inter-band relaxation dynamics.
In Fig.~\textcolor{magenta}{\ref{fig:8 E_p-1}} we show the on-site
polaron energy and band gap renormalization versus impurity-BEC coupling
constant. In both Fig.~\ref{fig:7 E_k}(a) and Fig.~\ref{fig:8 E_p-1}(a),
without loss of generality, we set the initial lower band on-site
energy $\varepsilon^{0}$ to zero. As shown in Fig.~\ref{fig:8 E_p-1}(b),
the band gap renormalization is almost temperature independent. These
quantities are only slightly affected by temperature due to different
variational transformation matrices $\Lambda_{\mathbf{\mathbf{q}}}$,
as predicted in Eq.~(\ref{eq:J_eff}). Although the band gap is not
significantly changed, this renormalization effect is important for
inter-band resonance conditions, which are required for Landau-Zener
tunneling to take place in a tilted lattice \cite{Parra-Murillo2013a,Parra-Murillo2014}.

\begin{figure}[h]
\includegraphics[bb=0bp 10bp 280bp 202bp,clip,width=0.5\textwidth]{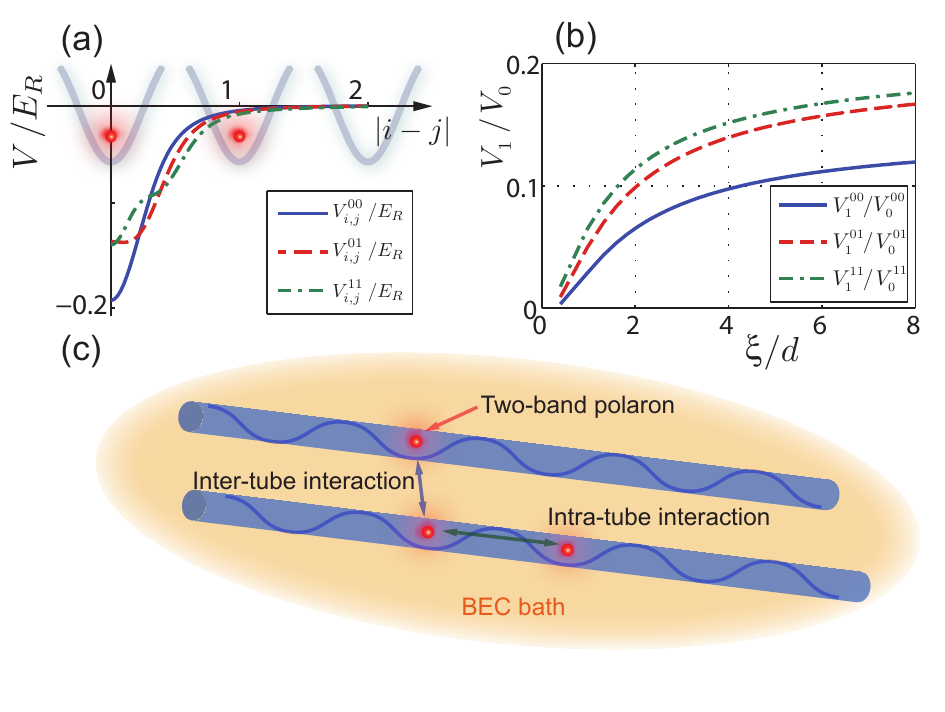}

\protect\caption{(a) Induced density-density interactions $V_{i,j}^{\alpha\beta}$
between different band at $\kappa=1$. (b) Ratio between induced on-site
and nearest neighbor interaction terms. The healing length $\xi$
is increased by reducing the condensate density $n_{0}$ from $1\times10^{14}\text{cm}^{-3}$
to $0.01\times10^{14}\text{cm}^{-3}$ while keeping the coupling constant
$\kappa=0.3\kappa_{c}$. (c) Effective polaronic interactions in a
multiple tube setup with two Bloch bands. Other parameters are $m_{I}=133,\,m_{B}=87$;
$V_{I}^{x}=9E_{R},\,V_{I}^{\perp}=25V_{I}^{x}$ and $a_{B}=100a_{0};\,n_{0}=1\times10^{14}\text{cm}^{-3}$.
\label{fig:9 effect_V_ij-1}}
\end{figure}

\subsection{Effective interactions between polarons}

Here we briefly discuss interactions between polarons, by considering
the additional interactions between impurities beyond the original
Hamiltonian (\ref{eq:Hamil_0}), which are given by 
\begin{equation}
\hat{V}_{0}=\sum_{i}\sum_{\alpha\beta}U^{\alpha\beta}\hat{n}_{i}^{\alpha}\left(\hat{n}_{i}^{\beta}-\delta_{\alpha\beta}\right),
\end{equation}
with the impurity on-site intra-/inter-band Hubbard-type interaction
$U^{\alpha\beta}$. After the polaron transformation, the coherent
Hamiltonian Eq.~(\ref{eq:coherent}) for the multi-polaron system
contains effective interactions $\hat{V}_{{\rm P}}$ with both intra-
and inter-band terms. In principle, the variational parameters $\lambda_{\mathbf{q}}$
should be determined by minimizing the total free energy. This is
hard for the many-impurity system. Here we assume that the polaron
parameters are not modified from the single polaron case, due to the
low density of impurities in our dilute system. The effective interaction
terms $\hat{V}_{{\rm P}}$ in Hamiltonian Eq.~(\ref{eq:coherent})
take the form:
\begin{alignat}{1}
\hat{V}_{{\rm P}}= & \sum_{i}\sum_{\alpha\beta}\left(U^{\alpha\beta}+V_{ii}^{\alpha\beta}\right)\hat{n}_{i}^{\alpha}\left(\hat{n}_{i}^{\beta}-\delta_{\alpha\beta}\right)\nonumber \\
+ & \sum_{i\neq j}\sum_{\alpha\beta}\left(V_{i,j}^{\alpha\beta}\hat{n}_{i}^{\alpha}\hat{n}_{j}^{\beta}+\sum_{\beta'\neq\beta}V_{i,j}^{\alpha;\beta\beta'}\hat{n}_{i}^{\alpha}\hat{a}_{j}^{\beta\dagger}\hat{a}_{j}^{\beta'}\right)\label{eq:interaction}
\end{alignat}
with long-range bath induced interaction terms:
\begin{alignat}{1}
V_{i,j}^{\alpha\beta}\equiv & -\sum_{\mathbf{q}}\hbar\omega_{\mathbf{q}}\cos\left[q_{x}(i-j)d\right]\nonumber \\
 & \:\quad\times\left(\Lambda_{\mathbf{q}}^{\alpha\alpha}\right)\left(2M_{\mathbf{q}}^{\beta\beta}-\Lambda_{\mathbf{q}}^{\beta\beta}\right)
\end{alignat}
and
\begin{alignat}{1}
V_{i,j}^{\alpha;\beta\beta'}\equiv & -\sum_{\mathbf{q}}\hbar\omega_{\mathbf{q}}\sin\left[q_{x}(i-j)d\right]\nonumber \\
 & \;\times\left(\Lambda_{\mathbf{q}}^{\alpha\alpha}\right)\left(2i\cdot M_{\mathbf{q}}^{\beta\beta'}\right)\langle\hat{K}_{j,j}^{\beta\beta'}\rangle_{T},
\end{alignat}
with
\begin{equation}
\hat{K}_{i,j}^{\alpha\beta}\equiv\left(\hat{\mathbf{X}}_{i}\right)_{\alpha\alpha}^{\dagger}\left(\hat{\mathbf{X}}_{j}\right)_{\beta\beta}=e^{-\sum_{\mathbf{q}}\left(\Lambda_{i,\mathbf{q}}^{\alpha\alpha}-\Lambda_{j,\mathbf{q}}^{\beta\beta}\right)^{*}\hat{b}_{\mathbf{q}}^{\dagger}-h.c.}.
\end{equation}

As it was shown in \cite{Bruderer2007,Santamore2011}, these induced
density-density interactions $V_{i,j}^{\alpha\beta}$ are always attractive
with $V_{i,j}^{\alpha\beta}<0$ and take the form of Yukawa-type interactions.
The first line in Eq.~(\ref{eq:interaction}) contains on-site effective
interactions between polarons, also within different bands. Due to
these attractive interactions $V_{i,j}^{\alpha\beta}$, the final
on-site interactions $\left(U^{\alpha\beta}+V_{ii}^{\alpha\beta}\right)$
must be repulsive for bosonic impurities in order to keep the system
stable. The second part in Eq.~(\ref{eq:interaction}) describes
the long-range part of these induced interactions. There are also
density-induced inter-band transitions, which are due to the inter-band
polaron-bath coupling.

In Fig.~\ref{fig:9 effect_V_ij-1}(a) we show the induced density-density
interactions $V_{i,j}^{\alpha\beta}$ versus distance $|i-j|$ between
polarons between each band. Since these interactions are induced by
the impurity-BEC coupling, the long-range behavior is related to the
condensate healing length $\xi$. Here we assume we can obtain large
values for $\xi$ by tuning the Bose-Bose scattering length $a_{B}$
to a small positive value or keeping the condensate density $n_{0}$
small. In Fig.~\ref{fig:9 effect_V_ij-1}(b) we also show the ratio
of $V_{i,j}^{\alpha\beta}$ between on-site and nearest neighbor terms
versus $\xi$. These ratios increase with the BEC healing length,
indicating a longer effective range of the interactions. The effects
of these interactions, which are beyond this paper, might include
new ordered polaron phases in our system with Hamiltonian Eq.~(\ref{eq:coherent}).
As shown in Fig.~\ref{fig:9 effect_V_ij-1}(c), if we consider that
the transverse confinement is due to a very deep optical lattice instead
of a single well, different 1D tubes could then interact due to the
bath-induced long range (attractive) interactions. A more rich phase
diagram is expected in such mixed-dimensional system coupled to a
bosonic bath \cite{Yin2011,Malatsetxebarria2013,Lan2014,Cai2014}.

\section{Lindblad equation and inter-band dynamics\label{sec:Lindblad-equation}}

The incoherent part of the Hamiltonian $\tilde{H}_{\text{inc}}=\tilde{H}-\langle\tilde{H}\rangle_{T}$
includes the residual coupling between polarons and phonon bath:\begin{widetext} 
\begin{alignat}{1}
\tilde{H}_{\text{inc}}= & -\sum_{\langle i,j\rangle}\sum_{\alpha}\left(J^{\alpha}\hat{T}_{i,j}^{\alpha\alpha}\right)\hat{a}_{i}^{\alpha\dagger}\hat{a}_{j}^{\alpha}+\sum_{i}\sum_{\alpha}\left[\sum_{\mathbf{q}}\hbar\omega_{\mathbf{q}}\hat{b}_{\mathbf{q}}^{\dagger}\left(M_{i,\mathbf{q}}^{\alpha\alpha}-\Lambda_{i,\mathbf{q}}^{\alpha\alpha}\right)^{*}+h.c.\right]\hat{a}_{i}^{\alpha\dagger}\hat{a}_{i}^{\alpha}\nonumber \\
+ & \sum_{i}\sum_{\alpha\neq\beta}\left\{ \sum_{\mathbf{q}}\hbar\omega_{\mathbf{q}}\left[\hat{b}_{\mathbf{q}}^{\dagger}\left(M_{i,\mathbf{q}}^{\alpha\beta}\right)^{*}\left(\hat{K}_{i,i}^{\alpha\beta}\right)-\left(\Lambda_{i,\mathbf{q}}^{\alpha\alpha}\right)\left(M_{i,\mathbf{q}}^{\alpha\beta}\right)^{*}\left(\hat{T}_{i,i}^{\alpha\beta}\right)\right]+h.c.\right\} \hat{a}_{i}^{\alpha\dagger}\hat{a}_{i}^{\beta}\nonumber \\
- & \sum_{i\neq j}\sum_{\alpha}\sum_{\alpha'\neq\beta'}\left\{ \sum_{\mathbf{q}}\hbar\omega_{\mathbf{q}}\left[\left(\Lambda_{i,\mathbf{q}}^{\alpha\alpha}\right)\left(M_{j,\mathbf{q}}^{\alpha'\beta'}\right)^{*}\left(\hat{T}_{j,j}^{\alpha'\beta'}\right)\right]+h.c.\right\} \hat{a}_{i}^{\alpha\dagger}\hat{a}_{i}^{\alpha}\hat{a}_{j}^{\alpha'\dagger}\hat{a}_{j}^{\beta'}.\label{eq:incoherent}
\end{alignat}
\end{widetext} with $\hat{T}_{i,j}^{\alpha\beta}\equiv\hat{K}_{i,j}^{\alpha\beta}-\langle\hat{K}_{i,j}^{\alpha\beta}\rangle_{T}$.
There are three different coupling terms in Eq.~(\ref{eq:incoherent}):
1) an intra-band part due to hopping of the polaron to nearest neighbor
sites and residual phonon dressing effects; 2) an inter-band part
due to polaron dynamics between two bands; and 3) a mixed many-body
term due to effective interactions between polarons.

The coherent part of the Hamiltonian Eq.~(\ref{eq:coherent}), as
calculated before, is a two-band Hubbard Hamiltonian. In order to
focus on inter-band relaxation effects, we restrict our investigations
to the case of a single polaron and ignore the interaction part $\hat{V}_{\text{P}}$.
The incoherent Hamiltonian part Eq.~(\ref{eq:incoherent}) can be
treated by a Lindblad master equation. After applying the Born-Markov
approximation, the Lindblad equation has the form:
\begin{equation}
\frac{d\rho\left(t\right)}{dt}=-i\left[\langle\tilde{H}\rangle_{T},\rho\left(t\right)\right]+\mathcal{L}_{I}\left[\rho\left(t\right)\right]
\end{equation}
with reduced density operator $\rho\left(t\right)$ of the polaron
system. All decoherence effects in the system are described by the
dissipator: 
\begin{alignat}{1}
\mathcal{L}_{I}\left[\rho\left(t\right)\right]= & \sum_{\alpha\beta}\sum_{\alpha'\beta'}\sum_{i,j}\sum_{i',j'}\gamma_{ij;i'j'}^{\alpha\beta;\alpha'\beta'}\nonumber \\
 & \times\left(C_{i'j'}^{\alpha'\beta'}\rho C_{ij}^{\alpha\beta\dagger}-\frac{1}{2}\left\{ C_{ij}^{\alpha\beta\dagger}C_{i'j'}^{\alpha'\beta'},\;\rho\right\} \right)\label{eq:Lindblad}
\end{alignat}
with the quantum jump operators $C_{ij}^{\alpha\beta}\equiv\left(\hat{a}_{i}^{\alpha\dagger}\hat{a}_{j}^{\beta}\right)$.
These jump operators describe the dynamics of polarons instead of
bare impurities, since they include the creation and annihilation
operators of the polaron. In the interaction picture with $\hat{O}\left(t\right)\equiv e^{-i\langle\tilde{H}\rangle_{T}t/\hbar}\hat{O}e^{i\langle\tilde{H}\rangle_{T}t/\hbar}$,
the decoherence rates are defined as:
\begin{equation}
\gamma_{ij;i'j'}^{\alpha\beta;\alpha'\beta'}\equiv2{\bf \text{Re}}\int_{0}^{\infty}d\tau e^{i\omega\tau}g_{ij;i'j'}^{\alpha\beta;\alpha'\beta'}\left(\tau\right)\label{eq:Lindblad-1}
\end{equation}
and the correlation functions $g_{ij;i'j'}^{\alpha\beta;\alpha'\beta'}\left(\tau\right)$
are defined in Appendix \ref{sub:App_Lindblad_polaron}. 

In this section, we focus our study on inter-band spontaneous relaxation
of the polaron. Due to longitudinal trapping, the energy scales for
intra- and inter-band dynamics are mismatched as $J^{0};J^{1}\ll\varepsilon_{{\rm P}}^{\Delta}$.
This allows us to use the rotating wave approximation (RWA) for inter-band
dynamics and decouple it from intra-band dynamics in Eq.~(\ref{eq:Lindblad_RWA}).
We then use the short-hand notation $\gamma_{i,j}^{01}\equiv\gamma_{ii;jj}^{01;01}$
and $C_{i}^{01}\equiv\left(\hat{a}_{i}^{0\dagger}\hat{a}_{i}^{1}\right)$
to describe the polaron relaxation\textbf{ }processes. Due to the
coupling with the bath, a polaron in the upper band can spontaneously
relax to the lower band with rate $\gamma_{i,j}^{01}$ and emit a
phonon. In our system, the corresponding phonon energy $\hbar\omega_{\mathbf{q}}$
for inter-band dynamics is much larger than the BEC temperature $k_{\text{B}}T$.
For inter-band dynamics, the phonon bath temperature is thus effectively
zero. In the interaction picture, this process is described by the
master equation
\[
\frac{d}{dt}\hat{\rho}\equiv\sum_{i,j}\gamma_{i,j}^{01}\left(C_{j}^{01}\hat{\rho}C_{i}^{01\dagger}-\frac{1}{2}\left\{ C_{i}^{01\dagger}C_{j}^{01},\;\hat{\rho}\right\} \right).
\]
The single polaron spontaneous relaxation rate $\gamma_{i,j}^{01}$
is calculated in Eq.~(\ref{eq:decay_rate_1}). It is convenient to
write the master equation for relaxation processes in momentum space,
as long as $\gamma_{i,j}^{01}$ only depends on the value of $\left(i-j\right)$:
\[
\frac{d}{dt}\hat{\rho}\equiv\sum_{q}\gamma_{q}\left(C_{q}^{01}\hat{\rho}C_{q}^{01\dagger}-\frac{1}{2}\left\{ C_{q}^{01\dagger}C_{q}^{01},\;\hat{\rho}\right\} \right),
\]
with $C_{q}^{01}\equiv\sum_{k}\left(\hat{a}_{k-q}^{0\dagger}\hat{a}_{k}^{1}\right)$
and $\hat{a}_{k}^{\alpha}$ is the polaron annihilation operator in
momentum space~\cite{Griessner2006,Griessner2007}. Here, $k,\,k-q$
are the quasi-momenta in the first Brillouin zone and $C_{q}\equiv C_{q+zG},\,z\in\mathbb{Z},\,G=2\pi/d$.
The relaxation rate is written as a function of longitudinal phonon
momentum:
\begin{equation}
\gamma_{q}=\frac{1}{N}\sum_{\left(i-j\right)}\gamma_{i,j}^{01}e^{-iq\left(i-j\right)d},\label{eq:gamma_q}
\end{equation}
where $N$ is lattice number. When $qd\gg2\pi$, the sum over with
$i\neq j$, involving polaron relaxation effects over different sites,
decays rapidly with $|i-j|$. 

The total relaxation rate $\gamma$ is a sum over longitudinal phonon
momenta, $\gamma\equiv\sum_{q}\gamma_{q}$ \cite{Griessner2007}.
From the expression of the correlation function $g_{ij}^{01}\left(\tau\right)$
in Eq.~(\ref{eq:correlation_g-1}), the polaronic inter-band relaxation
contains two terms. The first term in (\ref{eq:correlation_g-1})
is a single-phonon process, with the polaron absorbing (emitting)
one phonon from (to) the bath. This term is similar to Fermi's Golden
Rule except a renormalization factor $\langle\hat{K}_{i,i}^{10}\left(\tau\right)\hat{K}_{j,j}^{01}\left(0\right)\rangle_{T}$.
The second term describes higher order processes involving two phonons
begin absorbed (emitted), which are absent in Fermi's Golden Rule.
In the total relaxation rate, we will only consider the leading-order
single-phonon process in Eq.~(\ref{eq:decay_rate_1}) and neglect
the higher-order processes. At zero temperature, only spontaneous
emission of phonons is allowed with the single polaron spontaneous
relaxation rate
\begin{alignat}{1}
\gamma^{\textrm{P}}= & 2\text{Re}\int_{0}^{\infty}d\tau e^{i\varepsilon_{{\rm P}}^{\Delta}\tau/\hbar}\sum_{\mathbf{q}}\omega_{\mathbf{q}}^{2}|M_{\mathbf{q}}^{01}|^{2}e^{\text{-}i\omega_{\mathbf{q}}\tau}\nonumber \\
 & \qquad\times\langle\hat{K}_{i,i}^{10}\left(\tau\right)\hat{K}_{i,i}^{01}\left(0\right)\rangle_{T},\label{eq:gamma_polaron}
\end{alignat}
where $\langle\hat{K}_{i,i}^{10}\left(\tau\right)\hat{K}_{i,i}^{01}\left(0\right)\rangle_{T}$
is defined in Eq.~(\ref{eq:KiKj}) as a renormalization factor for
inter-band relaxation dynamics.

In order to investigate how polaron effects affect the inter-band
dynamics, we also derive a Lindblad equation for the original ``bare''
Hamiltonian in Eq.~(\ref{eq:Hamil_0}), before the Lang-Firsov transformation
had been applied. The corresponding single-impurity spontaneous relaxation
rate in this case is 
\begin{equation}
\gamma^{0}=2\text{Re}\int_{0}^{\infty}d\tau e^{i\varepsilon^{\Delta}\tau/\hbar}\sum_{\mathbf{q}}\omega_{\mathbf{q}}^{2}|M_{\mathbf{q}}^{01}|^{2}e^{\text{-}i\omega_{\mathbf{q}}\tau},\label{eq:gamma_bare}
\end{equation}
and reduces to Fermi's Golden Rule formula:
\begin{equation}
\gamma^{0}=2\pi\sum_{\mathbf{q}}\omega_{\mathbf{q}}^{2}|M_{\mathbf{q}}^{01}|^{2}\delta\left(\hbar\omega_{\mathbf{q}}-\varepsilon^{\Delta}\right).\label{eq:golden_rule}
\end{equation}
Comparing Eq.~(\ref{eq:gamma_polaron}) and Eq.~(\ref{eq:gamma_bare}),
we observe that the renormalization of the band gap from $\varepsilon^{\Delta}$
to $\varepsilon_{{\rm P}}^{\Delta}$ increases the energy of the phonon
which is created, due to energy conservation. The polaron relaxation
rate $\gamma^{\textrm{P}}$ is also reduced by an additional exponential
factor, since the coupling between impurity and bath is also renormalized. 

Before we discuss the numerical results for the relaxation rate, let
us first look at the behavior of $\gamma_{q}$ as a function of phonon
momentum in the longitudinal direction. For the bare particle relaxation
rate, the prediction of Fermi's golden rule gives:
\begin{alignat}{1}
\gamma_{q}^{0}\equiv & 2\text{Re}\int_{0}^{\infty}d\tau e^{i\varepsilon^{\Delta}\tau/\hbar}\sum_{q_{y},q_{z}}\omega_{\mathbf{q}}^{2}|M_{\mathbf{\mathbf{q}}}^{01}|^{2}e^{\text{-}i\omega_{\mathbf{q}}\tau}\nonumber \\
\approx & 2\pi\sum_{q_{\perp}}\omega_{\mathbf{q}}^{2}|M_{\mathbf{q}}^{01}|^{2}\delta\left(\hbar\omega_{\mathbf{q}}-\varepsilon^{\Delta}\right).\label{eq:gamma_bare_q}
\end{alignat}
In a deep lattice, the impurity-phonon coupling matrices $M_{\mathbf{q}}^{\alpha\beta}$
can be approximated by Gaussian functions in Eq.~(\ref{eq:coupling_factor}).
Then the value of Fermi's golden rule for $\gamma_{q}^{0}$ in Eq.~(\ref{eq:gamma_bare_q})
is found to be:
\begin{alignat}{1}
\gamma_{q}^{0}\approx & \left(\frac{m_{B}}{\hbar^{2}}\right)\frac{n_{0}}{L_{x}}g_{IB}^{2}\nonumber \\
 & \times\left(q^{2}\sigma_{x}^{2}/2\right)e^{-\left(q^{2}\sigma_{x}^{2}+q_{\perp}^{2}\sigma_{\perp}^{2}\right)/2}\label{eq:gamma_bare_q_app}
\end{alignat}
where $L_{x}$ is the phonon quantization length (analogous to the
quantization volume $\Omega$) in the longitudinal direction. The
phonon momentum in the transverse direction $q_{\perp}$ is fixed
by energy conservation $\hbar\omega_{\mathbf{q}}=\varepsilon^{\Delta}\approx\hbar\omega_{x}$,
where $|\mathbf{q}|^{2}=q^{2}+q_{\perp}^{2}$ and $\omega_{x}$ is
the longitudinal oscillation frequency. An additional consequence
of energy conservation, is that $\gamma_{q}$ is cut off when $\hbar\omega_{\mathbf{q}}\left(|\mathbf{q}|=q\right)=\hbar\omega_{x}$.
The cut-off is 
\begin{equation}
|q\sigma_{x}|\approx\sqrt{2\frac{m_{B}}{m_{I}}},\label{eq:cut_off}
\end{equation}
which only depends on the mass ratio $m_{I}/m_{B}$ between impurity
and BEC particles. 

On the other hand, the polaron relaxation rate $\gamma_{q}^{\textrm{P}}$
versus longitudinal phonon momentum needs to be calculated numerically
from Eq.~(\ref{eq:gamma_q}) and Eq.~(\ref{eq:decay_rate_1}). As
we discussed before for the total relaxation rate Eq.~(\ref{eq:gamma_polaron}),
we consider only the single-phonon emitting processes and calculate
$\gamma_{q}^{\textrm{P}}$ as:
\begin{alignat}{1}
\gamma_{q}^{\textrm{P}}\approx & \,2\text{Re}\int_{0}^{\infty}d\tau e^{i\varepsilon_{{\rm P}}^{\Delta}\tau/\hbar}\sum_{q_{y},q_{z}}\omega_{\mathbf{q}}^{2}|M_{\mathbf{q}}^{01}|^{2}e^{-i\omega_{\mathbf{q}}\tau}\nonumber \\
 & \qquad\cdot\langle\hat{K}_{i,i}^{10}\left(\tau\right)\hat{K}_{i,i}^{01}\left(0\right)\rangle_{T}.\label{eq:gamma_polaron_q}
\end{alignat}
Comparing with Eq.~(\ref{eq:gamma_bare_q}), the polaron relaxation
rate has a similar behavior except for an additional renormalization
factor. Due to energy conservation, $\gamma_{q}^{\textrm{P}}$ also
has a cut-off when $\hbar\omega_{\mathbf{q}}\left(|\mathbf{q}|=q\right)=\varepsilon_{{\rm P}}^{\Delta}$.
The renormalized polaron band gap $\varepsilon_{{\rm P}}^{\Delta}$,
which is larger than the bare gap $\varepsilon^{\Delta}$, will shift
this cut-off position to higher phonon momentum. At the same time,
the renormalization factor in Eq.~(\ref{eq:gamma_polaron_q}) will
reduce the value of $\gamma_{q}^{\textrm{P}}$.

\begin{figure}
\includegraphics[bb=0bp 0bp 305bp 213bp,clip,width=0.5\textwidth]{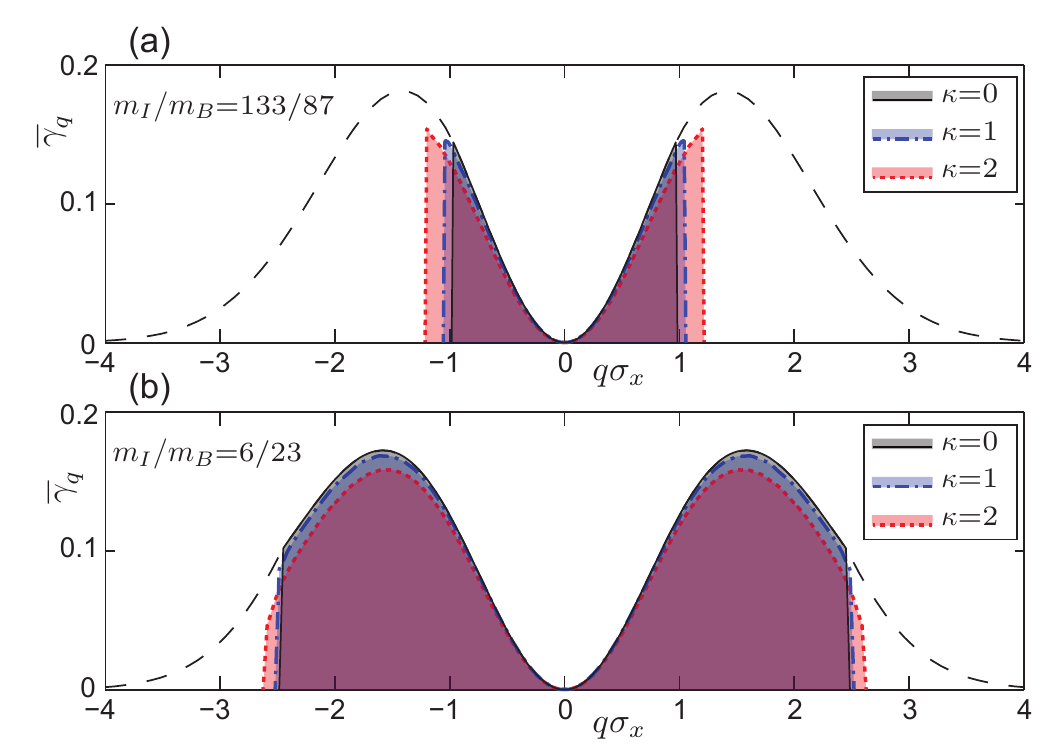}

\protect\caption{Normalized relaxation rate $\gamma_{q}^{\textrm{P}}$ and $\gamma_{q}^{0}$
as functions of longitudinal momentum $q\sigma_{x}$ with different
impurity-BEC mass ratio. The dashed black lines indicate Fermi's Golden
Rule results $\gamma_{q}^{0}$ from Eq.~(\ref{eq:gamma_bare_q_app})
with momentum cut-off given by the solid black line ($\kappa=0$).
The blue (red) curves indicate normalized polaron relaxation rate
$\gamma_{q}^{\textrm{P}}$ at coupling $\kappa=1\,(2)$. The total
relaxation rates $\gamma=\sum_{q}\gamma_{q}$ are shown as the area
below the corresponding curves. (a) System with $m_{I}/m_{B}=133/87$,
i.e. $^{133}\text{Cs}$ impurity in $^{87}\text{Rb}$ BEC. (b) System
with $m_{I}/m_{B}=6/23$, i.e. $^{6}\text{Li}$ impurity in $^{23}\text{Na}$
BEC. Other parameters are $V_{I}^{x}=9E_{R},\,V_{I}^{\perp}=25V_{I}^{x}$
and $a_{B}=100a_{0},\,n_{0}=1\times10^{14}\text{cm}^{-3}$. \label{fig:decay-q}}
\end{figure}

In Fig.~\ref{fig:decay-q} we show the behavior of the relaxation
rate $\gamma_{q}^{\textrm{P}}$ and $\gamma_{q}^{0}$ as functions
of longitudinal momentum $q\sigma_{x}$ with different impurity-BEC
mass ratio. In order to compare the differences between $\gamma_{q}^{\textrm{P}}$
and $\gamma_{q}^{0}$ at different coupling, we divide both $\gamma_{q}^{\textrm{P}}$
and $\gamma_{q}^{0}$ by $\kappa^{2}$ and then normalize them. In
this way, the Fermi's golden rule value $\gamma_{q}^{0}$ always stays
the same for different coupling constants $\kappa$. As shown in Fig.~\ref{fig:decay-q},
the dashed black lines indicate normalized $\gamma_{q}^{0}$ from
Eq.~(\ref{eq:gamma_bare_q_app}), while the cut-off of $|q\sigma_{x}|$
in Eq.~(\ref{eq:cut_off}) is indicated by black lines. The total
relaxation rate $\gamma^{0}=\sum_{q}\gamma_{q}^{0}$ from Fermi's
golden rule is the gray area below the black curve. At $\kappa=0$,
the polaronic relaxation rate $\gamma_{q}^{\textrm{P}}$ is identical
to the results from the Golden Rule.

On the other hand, as the impurity-BEC coupling is increased, the
polaron relaxation rate $\gamma_{q}^{\textrm{P}}$ from Eq.~(\ref{eq:gamma_polaron_q})
is renormalized by the polaron band gap $\varepsilon_{{\rm P}}^{\Delta}$
and renormalization factor. The increased polaron band gap will involve
more phonons by shifting the cut-off momentum (\ref{eq:cut_off}),
while the renormalization factor reduces the whole momentum range.
In Fig.~\ref{fig:decay-q} we also show the polaron relaxation rate
$\gamma_{q}^{\textrm{P}}$ with blue (red) curve at $\kappa=1(2)$
and total relaxation rate $\gamma^{\textrm{P}}$ with blue (red) area.

As shown in Fig.~\ref{fig:decay-q}, the renormalization of the relaxation
rate is different for various impurity-BEC mass ratios. For larger
mass ratio (heavy impurity) such as a system with single $^{133}\text{Cs}$
impurity coupled with $^{87}\text{Rb}$ BEC, the polaron relaxation
processes will be enhanced by the shift of the cut-off momentum and
increase the total rate. On the other hand, for smaller mass ratio
(light impurity) such as a $^{6}\text{Li}$ impurity coupled with
$^{23}\text{Na}$ BEC, the higher momentum cut-off for $|q\sigma_{x}|$
is not so important due to Gaussian decay of $\gamma_{q}$. In this
case, the renormalization factor will reduce $\gamma_{q}^{\textrm{P}}$
as well as the total relaxation rate $\gamma^{\textrm{P}}$. We can
also expect that, for extremely strong impurity-BEC coupling, the
total relaxation rate will be reduced due to this renormalization
factor with any mass ratio. 

\begin{figure}
\includegraphics[bb=0bp 0bp 320bp 140bp,clip,width=0.5\textwidth]{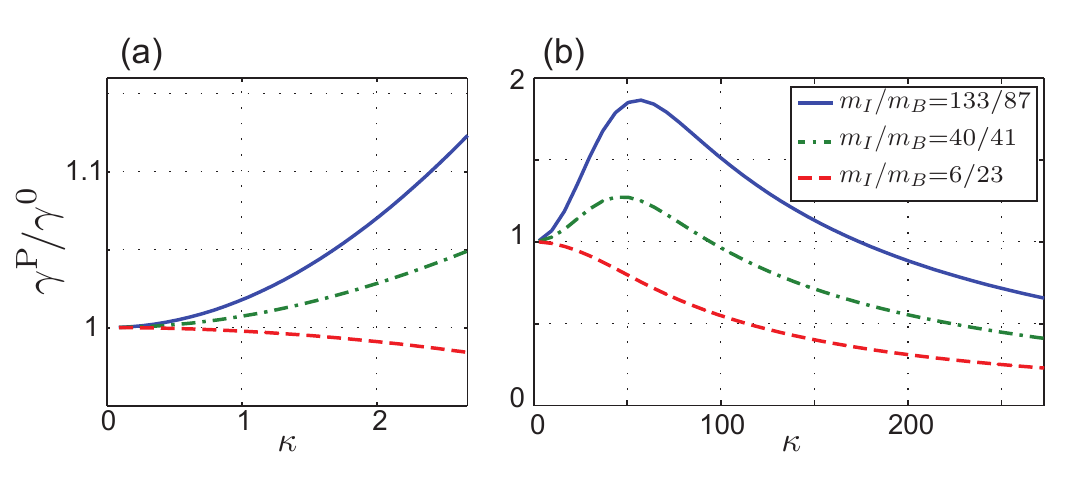}

\protect\caption{Ratio between polaron inter-band relaxation rate $\gamma^{\text{P}}$
and Fermi Golden Rule results $\gamma^{0}$ at different impurity-BEC
mass ratio. (a) In the weak coupling regime, this ratio is increased
with the coupling constant $\kappa$ for a heavy impurity coupled
to a light BEC bath, and decreased for a light impurity coupled to
a heavy BEC bath. The boson-boson scattering length here is chosen
as $a_{B}=100a_{0}$ and $\kappa_{c}=2.4$. (b) In the strong coupling
regime, the polaron inter-band relaxation processes for any impurity-BEC
mass ratio are suppressed. This leads to an inter-band self-trapping
effect. The boson-boson scattering length here is $a_{B}=0.2a_{0}$
and $\kappa_{c}=255$. We use $m_{I}=133,\,m_{B}=87$ for a $^{133}\text{Cs}$
impurity and $^{87}\text{Rb}$ BEC (blue line), $m_{I}=40,\,m_{B}=41$
for a $^{40}\text{K}$ impurity and $^{41}\text{K}$ BEC (green line),
and $m_{I}=6,\,m_{B}=23$ for a $^{6}\text{Li}$ impurity and $^{23}\text{Na}$
BEC (red line). Other parameters are $V_{I}^{x}=9E_{R},\,V_{I}^{\perp}=25V_{I}^{x}$
and $n_{0}=1\times10^{14}\text{cm}^{-3}$.\label{fig_decay}}
\end{figure}

Finally we compare the polaronic inter-band relaxation rate $\gamma^{\text{P}}$
and Fermi Golden Rule results $\gamma^{0}$ in Fig.$\:$\ref{fig_decay}
as a function of coupling constant and for different impurity-BEC
mass ratios. As we have already seen in Fig.$\:$\ref{fig:decay-q},
the polaron formation will renormalize the inter-band relaxation rate
differently, depending on the mass ratio. 

We first consider the weak impurity-BEC coupling regime with small
$\kappa$ in Fig.$\:$\ref{fig_decay}(a). For a heavy impurity coupled
to a light BEC bath, the inter-band relaxation process involves more
phonon modes and will be enhanced. The difference in the ratio $\gamma^{\text{P}}/\gamma^{0}$
between polaronic relaxation and the Fermi Golden Rule result increases
with coupling constant $\kappa$. On the other hand, for a light impurity
coupled to a heavy BEC, the impurity is dressed by heavy phonons in
each band and tends to localize in the same band. Although the inter-band
relaxation process does involve more phonon modes than the Fermi Golden
Rule result, this effect is highly suppressed due to Gaussian decay
of $\gamma_{q}$ with high momentum. The ratio $\gamma^{\text{P}}/\gamma^{0}$
will be reduced in this system. We also notice that the ratio $\gamma^{\text{P}}/\gamma^{0}$
is not sensitive to the longitudinal trapping potential.

On the other hand, we can access the strong coupling regime by tuning
the Bose-Bose scattering $a_{B}$ to a small positive value, since
the allowed maximum coupling constant in Eq.~(\ref{eq:kappa_C})
goes as $\kappa_{c}\propto1/\left(n_{0}a_{B}^{3}\right)^{1/4}$. For
the strong impurity-BEC coupling region with large $\kappa$ in Fig.$\:$\ref{fig_decay}(b),
though the Lindblad master equation might not be accurate enough for
such strong inter-band coupling, we can still obtain some qualitative
insight. In each band, the polaron is tightly dressed by phonons with
different coupling strength, such that the polaron behaves as a quasi-particle
with rather different properties in each band. Also, the band gap
$\varepsilon_{{\rm P}}^{\Delta}$ is enhanced in comparison to the
bare case, so the polaron cannot hop between bands by creating or
annihilating phonons. As shown in Fig.$\:$\ref{fig_decay}(b), we
indeed obtain a suppressed inter-band relaxation process. This \textit{inter-band
self-trapping }effect is expected in a strongly coupled impurity-BEC
system. In a realistic impurity-BEC system, this effect might be also
observed together with the well known \textit{self-trapping} effect
due to deformation of BEC \cite{Lee1992,Cucchietti2006,Kalas2006a,Luhmann2008,Bruderer2008c,Roberts2009,Heinze2011}.

\section{Conclusions\label{sec:Conclusion}}

In conclusion, we have studied a two-band Hamiltonian with Fr\"{o}hlich
impurity-phonon coupling with both intra- and inter-band terms. Such
a Hamiltonian can be realized in experiments where few impurities
are immersed in a Bose-Einstein condensate of another species. The
impurities are trapped by an anisotropic optical lattice behave as
quasi-1D particles. Based on the Lang-Firsov transformation, we have
derived and applied a variational two-band polaron transformation.
We have calculated the coherent part of the resulting effective Hamiltonian
with two (polaron) bands. In each band the impurity is dressed by
phonons as a quasi-particle (polaron) with different properties. The
polaronic intra-band coherent transport and polaron effective mass
are both renormalized. Due to the coupling with bath, there are also
induced on-site polaron energy shifts, and long-range interactions
between different polarons. 

In order to account for the residual incoherent coupling between polaron
and bath, we have derived a Lindblad master equation, and focused
on the single-polaron inter-band relaxation dynamics. Comparing to
Fermi's Golden Rule calculations for the bare impurities, we found
the renormalization of the relaxation rate to depend on the mass ratio
of impurity and BEC particles. These polaronic effects in the inter-band
relaxation dynamics might be observed in ongoing experiments. In the
strong coupling limit of the two-band Fr\"{o}hlich Hamiltonian, the
impurity is \textit{inter-band self-trapped} and can not tunnel between
bands by creating or annihilating phonons. 

~
\begin{acknowledgments}
The authors thank A. Daley, E. Demler, R. Gerritsma, F. Grusdt, G.
Pupillo, T. Rentrop, R. Schmidt, T. Shi and A. Widera for fruitful
discussions. Support by the Deutsche Forschungsgemeinschaft (DFG)
via Sonderforschungsbereich SFB/TR 49, Forschergruppe FOR 801 and
the high-performance computing center LOEWE-CSC is gratefully acknowledged. 
\end{acknowledgments}

\appendix%\appendixpage

\onecolumngrid

%\begin{widetext} 

\addcontentsline{toc}{section}{Appendices}

%\markboth{APPENDICES}{} 

\section{Two-band transformation and coherent Hamiltonian\label{sub:App_Polaron-transformation} }

\subsection{Two-band polaron transformation}

In this section, we derive the two-band polaron transformation in
detail. The transformation operator $\hat{S}$ in Eq.$\:$(\ref{eq:transformation_S})
can be written as $\hat{S}=\sum_{i,\alpha,\beta}\hat{C}_{i}^{\alpha,\beta}\hat{a}_{i}^{\alpha\dagger}\hat{a}_{i}^{\beta}$
with $\hat{C}_{i}^{\alpha,\beta}\equiv\sum_{\mathbf{q}}\Lambda_{\mathbf{q}}^{\alpha,\beta}e^{i\mathbf{q}\cdot\mathbf{R}_{i}}\left(\hat{b}_{-\mathbf{q}}^{\dagger}-\hat{b}_{\mathbf{q}}\right)$
or in matrix form as $\hat{\mathbf{C}}_{i}\equiv\sum_{\mathbf{q}}\mathbf{\Lambda}_{\mathbf{q}}\left(\hat{\mathbf{b}}_{-\mathbf{q}}^{\dagger}-\hat{\mathbf{b}}_{\mathbf{q}}\right)e^{i\mathbf{q}\cdot\mathbf{R}_{i}}$.
By using the relation $\left[\hat{a}_{i}^{\alpha\dagger}\hat{a}_{i}^{\beta},\;\hat{a}_{j}^{\gamma}\right]=-\delta_{ij}\delta_{\alpha\gamma}\hat{a}_{i}^{\beta}$
, which is valid for both bosons and fermions, we obtain the transformed
impurity annihilation operator as:

\begin{alignat}{1}
\left[\hat{S},\;\hat{a}_{i}^{\alpha}\right]= & -\sum_{\beta}\left(\hat{\mathbf{C}}_{i}\right)_{\alpha\beta}\hat{a}_{i}^{\beta};\quad\left[\hat{S},\;\left[\hat{S},\;\hat{a}_{i}^{\alpha}\right]\right]=+\sum_{\beta}\left(\hat{\mathbf{C}}_{i}\hat{\mathbf{C}_{i}}\right)_{\alpha\beta}\hat{a}_{i}^{\beta};\quad\cdots;\\
e^{\hat{S}}\hat{a}_{i}^{\alpha}e^{-\hat{S}}= & \sum_{\beta}\left(e^{-\hat{\mathbf{C}}_{i}}\right)_{\alpha\beta}\hat{a}_{i}^{\beta}=\sum_{\beta}\left(\hat{\mathbf{X}}_{i}\right)_{\alpha\beta}\hat{a}_{i}^{\beta},
\end{alignat}
with $\hat{\mathbf{X}}_{i}=e^{-\hat{\mathbf{C}_{i}}}$ is defined
in Eq.~(\ref{eq:transform_a}). The phonon annihilation operators
can also be transformed in a similar way as:
\begin{alignat}{1}
\left[\hat{S},\;\hat{b}_{\mathbf{q}}\right]= & \sum_{i,\alpha,\beta}\left(\left[\hat{\mathbf{C}}_{i},\;\hat{\mathbf{b}}_{\mathbf{q}}\right]\right)_{\alpha,\beta}\hat{a}_{i}^{\alpha\dagger}\hat{a}_{i}^{\beta};\\
\left[\hat{S},\;\left[\hat{S},\;\hat{b}_{\mathbf{q}}\right]\right]= & \left[\sum_{i,\alpha,\beta}\hat{C}_{i}^{\alpha,\beta}\hat{a}_{i}^{\alpha\dagger}\hat{a}_{i}^{\beta},\;\sum_{i',\alpha',\beta'}\left(\left[\hat{\mathbf{C}}_{i'},\hat{\mathbf{b}}_{\mathbf{q}}\right]\right)_{\alpha',\beta'}\hat{a}_{i'}^{\alpha'\dagger}\hat{a}_{i'}^{\beta'}\right]=\sum_{i,\alpha,\beta}\left(\left[\hat{\mathbf{C}}_{i},\;\left[\hat{\mathbf{C}}_{i},\hat{\mathbf{b}}_{\mathbf{q}}\right]\right]\right)_{\alpha\beta}\hat{a}_{i}^{\alpha\dagger}\hat{a}_{i}^{\beta};\\
\vdots\nonumber 
\end{alignat}
and thus
\begin{alignat}{1}
e^{\hat{S}}\hat{b}_{\mathbf{q}}e^{-\hat{S}}=\: & \hat{b}_{\mathbf{q}}+\sum_{i,\alpha,\beta}\left(e^{\hat{\mathbf{C}}_{i}}\hat{\mathbf{b}}_{\mathbf{q}}e^{-\hat{\mathbf{C}}_{i}}-\hat{\mathbf{b}}_{\mathbf{q}}\right)_{\alpha\beta}\hat{a}_{i}^{\alpha\dagger}\hat{a}_{i}^{\beta}=\hat{b}_{\mathbf{q}}+\sum_{i,\alpha,\beta}\left(\hat{\mathbf{X}}_{i}^{\dagger}\hat{\mathbf{b}}_{\mathbf{q}}\hat{\mathbf{X}}_{i}-\hat{b}_{\mathbf{q}}\right)_{\alpha\beta}\hat{a}_{i}^{\alpha\dagger}\hat{a}_{i}^{\beta}.\label{eq:trans_b}
\end{alignat}
Different from the single band case, all higher commutators in Eq.~(\ref{eq:trans_b})
are nonzero except when $\left[\mathbf{\Lambda}_{\mathbf{q}},\:\mathbf{\Lambda}_{\mathbf{\mathbf{q}'}}\right]=0$.
The Hamiltonian is transformed as: 
\begin{alignat}{1}
\tilde{H}= & -\sum_{\langle i,j\rangle}\sum_{\alpha\beta}\left(\hat{\mathbf{X}}_{i}^{\dagger}\mathbf{J}\hat{\mathbf{X}}_{j}\right)_{\alpha\beta}\hat{a}_{i}^{\alpha\dagger}\hat{a}_{j}^{\beta}+\sum_{i}\sum_{\alpha\beta}\left(\hat{\mathbf{X}}_{i}^{\dagger}\mathbf{\varepsilon}\hat{\mathbf{X}}_{i}\right)_{\alpha\beta}\hat{a}_{i}^{\alpha\dagger}\hat{a}_{i}^{\beta}\nonumber \\
+ & \sum_{\mathbf{q}}\hbar\omega_{\mathbf{q}}\left(\hat{b}_{\mathbf{q}}^{\dagger}+\sum_{i,\alpha,\beta}\left(\hat{\mathbf{X}}_{i}^{\dagger}\hat{\mathbf{b}}_{\mathbf{q}}^{\dagger}\hat{\mathbf{X}}_{i}-\hat{\mathbf{b}}_{\mathbf{q}}^{\dagger}\right)_{\alpha\beta}\hat{a}_{i}^{\alpha\dagger}\hat{a}_{i}^{\beta}\right)\left(\hat{b}_{\mathbf{q}}+\sum_{j,\alpha',\beta'}\left(\hat{\mathbf{X}}_{j}^{\dagger}\hat{\mathbf{b}}_{\mathbf{q}}\hat{\mathbf{X}}_{j}-\hat{\mathbf{b}}_{\mathbf{q}}\right)_{\alpha'\beta'}\hat{a}_{j}^{\alpha'\dagger}\hat{a}_{j}^{\beta'}\right)\nonumber \\
+ & \sum_{i,\alpha,\beta}\sum_{\mathbf{q}}\hbar\omega_{\mathbf{q}}\left(\hat{b}_{-\mathbf{q}}^{\dagger}+\sum_{j,\alpha',\beta'}\left(\hat{\mathbf{X}}_{j}^{\dagger}\hat{\mathbf{b}}_{-\mathbf{q}}^{\dagger}\hat{\mathbf{X}}_{j}-\hat{\mathbf{b}}_{-\mathbf{q}}^{\dagger}\right)_{\alpha'\beta'}\hat{a}_{j}^{\alpha'\dagger}\hat{a}_{j}^{\beta'}+\hat{b}_{\mathbf{q}}+\sum_{j,\alpha',\beta'}\left(\hat{\mathbf{X}}_{j}^{\dagger}\hat{\mathbf{b}}_{\mathbf{q}}\hat{\mathbf{X}}_{j}-\hat{\mathbf{b}}_{\mathbf{q}}\right)_{\alpha'\beta'}\hat{a}_{j}^{\alpha'\dagger}\hat{a}_{j}^{\beta'}\right)\nonumber \\
 & \qquad\qquad\cdot\left(\hat{\mathbf{X}}_{i}^{\dagger}\mathbf{M}_{i,\mathbf{q}}\hat{\mathbf{X}}_{i}\right)_{\alpha\beta}\hat{a}_{i}^{\alpha\dagger}\hat{a}_{i}^{\beta}.
\end{alignat}
 The resulting polaron Hamiltonian is presented in Eq.~(\ref{eq:transformed-Hamil_0}).

\subsection{Coherent Hamiltonian under thermal average\label{sub:App_Thermal-Average-calculation}}

For completeness, we describe a general calculation of coherent Hamiltonian
after two-band transformation. We apply a \textit{shorthand notation}
technique, which was used by Sibley and Munn \cite{Munn1985a,Munn1985b},
to our two-band system. The transformed Hamiltonian in Eq.~(\ref{eq:Hamil_0})
is complicated since it has no assumptions for the variational parameters
$\mathbf{\mathbf{\Lambda}}_{\mathbf{q}}$ except the symmetry relations
in Eq.~(\ref{eq:M_relation}). On the other hand, for our specific
choice of assuming diagonal matrices for the variational parameters,
all of the expressions in this section can be shown to be straightforward
and simple.

The coherent part $\langle\tilde{H}\rangle_{T}$ of Eq.~(\ref{eq:Hamil_0})
can be calculated by averaging over the phonon bath. Here we assume
the phonon bath is thermally distributed at the BEC temperature. From
Eq.~(\ref{eq:Hamil_0}) we find these calculations require several
types of thermal average values such as (I) $\langle\left(\hat{\mathbf{X}}_{i}^{\dagger}\mathbf{J}\hat{\mathbf{X}}_{j}\right)\rangle_{T}$,
(II) $\langle\left(\hat{\mathbf{X}}_{i}^{\dagger}\hat{\mathbf{b}}_{\mathbf{q}}^{\dagger}\mathbf{M}_{i,\mathbf{q}}^{\dagger}\hat{\mathbf{X}}_{i}\right)\rangle_{T}$,
(III) $\langle\left(\hat{\mathbf{X}}_{i}^{\dagger}\hat{\mathbf{b}}_{\mathbf{q}}^{\dagger}\hat{\mathbf{b}}_{\mathbf{q}}\hat{\mathbf{X}}_{i}\right)\rangle_{T}$
and (IV) $\langle\left(\hat{\mathbf{X}}_{i}^{\dagger}\hat{\mathbf{b}}_{\mathbf{q}}^{\dagger}\hat{\mathbf{X}}_{i}-\hat{\mathbf{b}}_{\mathbf{q}}^{\dagger}\right)_{\alpha\beta}\left(\hat{\mathbf{X}}_{i}^{\dagger}\hat{\mathbf{b}}_{\mathbf{q}}\hat{\mathbf{X}}_{i}-\hat{\mathbf{b}}_{\mathbf{q}}\right)_{\alpha'\beta'}\rangle_{T}$.
Note that some terms inside $\langle\left(\cdots\right)\rangle_{T}$
are $2\times2$ matrices. Here we will discuss these terms in details,
without making any assumption for the parameters $\mathbf{\Lambda}_{\mathbf{q}}$
except that they obey the same symmetry relations as for $\mathbf{M}_{\mathbf{q}}$
as given in Eq.~(\ref{eq:M_relation}).

\subsubsection*{Part I: $\langle\left(\hat{\mathbf{X}}_{i}^{\dagger}\mathbf{J}\hat{\mathbf{X}}_{j}\right)\rangle_{T}$
and $\langle\left(\hat{\mathbf{X}}_{i}^{\dagger}\mathbf{\varepsilon}\hat{\mathbf{X}}_{j}\right)\rangle_{T}$ }

The first coherent part describes a matrix $\mathbf{J}$ or $\mathbf{\varepsilon}$
transformed by operator $\hat{\mathbf{X}}_{i}$. We use a general
$2\times2$ matrix $\mathcal{F}$ instead of $J$ or $\varepsilon$
as:
\begin{equation}
\langle\left(\hat{\mathbf{X}}_{i}^{\dagger}\mathcal{F}\hat{\mathbf{X}}_{j}\right)_{\alpha\beta}\rangle_{T}=\sum_{\alpha'\beta'}\left(\mathcal{F}\right)_{\alpha'\beta'}\langle\left(\hat{\mathbf{X}}_{i}^{\dagger}\right)_{\alpha\alpha'}\left(\hat{\mathbf{X}}_{j}\right)_{\beta'\beta}\rangle_{T}\equiv\sum_{\alpha'\beta'}\left(\mathcal{F}\right)_{\alpha'\beta'}\langle\left(e^{\hat{\mathbf{C}}_{i}-\hat{\mathbf{C}}_{j}}\right)_{\alpha\alpha';\beta'\beta}\rangle_{T}.\label{eq:first-shorthand}
\end{equation}
In the last step, we apply the shorthand notation technique and imply
index $\alpha,\alpha'$ for all functions of $\hat{\mathbf{C}}_{i}$
and the index $\beta',\beta$ for all functions of $\hat{\mathbf{C}}_{j}$
\cite{Munn1985a,Stojanovic2004}. This shorthand notation ignores
the commutation of matrices and collects all terms of $\hat{\mathbf{C}}_{i}$
together and all $\hat{\mathbf{C}}_{j}$ together, before applying
the indices. For an arbitrary operator $\hat{\mathbf{C}}$ and an
arbitrary type of averaging involved, the expansion up to the second-order
cumulant reads 
\begin{equation}
\langle\exp\hat{\mathbf{C}}\rangle_{T}\simeq\exp\left\{ \langle\hat{\mathbf{C}}\rangle_{T}+\frac{1}{2}\left(\langle\hat{\mathbf{C}}^{2}\rangle_{T}-\langle\hat{\mathbf{C}}\rangle_{T}^{2}\right)\right\} .
\end{equation}
This expansion is exact if applied without truncation directly to
equation (\ref{eq:first-shorthand}) when the matrices $\hat{\mathbf{C}}$
commute, i.e. $[\hat{\mathbf{C}}_{i},\hat{\mathbf{C}}_{j}]=0$. However,
we make the approximation~\cite{Munn1985a,Stojanovic2004}, that
the exponential in the short-hand notation also follows such a cumulant
expansion. For small deviations from community, this will be a negligible
approximation. Since the operators $\hat{\mathbf{C}}_{i}=\sum_{\mathbf{q}}\mathbf{\Lambda}_{\mathbf{q}}e^{i\mathbf{q}\cdot\mathbf{R}_{i}}\left(\hat{\mathbf{b}}_{-\mathbf{q}}^{\dagger}-\hat{\mathbf{b}}_{\mathbf{q}}\right)$
contain only creation and annihilation operators, the coherent part
$\langle\hat{\mathbf{C}}\rangle_{T}=0$. In this case, we have $\langle\exp\hat{\mathbf{C}}\rangle_{T}\simeq\exp\left(\langle\hat{\mathbf{C}}^{2}\rangle_{T}/2\right)$. 

For the single band calculation, this relation reduces to the \textit{Bloch
identity} and is valid exactly \cite{Grosso2013}. However, in the
two band system this relation is only an approximation because higher
order terms are also present. The accuracy depends on the commutator
$\left[\mathbf{\Lambda}_{\mathbf{q}},\;\mathbf{\Lambda}_{\mathbf{q}'}\right]$.
Here we first apply this approximation with general parameters $\mathbf{\Lambda}_{\mathbf{q}}$:
\begin{alignat}{1}
 & \langle\left(e^{\hat{\mathbf{C}}_{i}-\hat{\mathbf{C}}_{j}}\right)_{\alpha\alpha';\beta'\beta}\rangle_{T}\simeq\left(e^{\frac{1}{2}\langle\left(\hat{\mathbf{C}}_{i}-\hat{\mathbf{C}}_{j}\right)^{2}\rangle_{T}}\right)_{\alpha\alpha';\beta'\beta}=\left(e^{\frac{1}{2}\langle\hat{\mathbf{C}}_{i}^{2}\rangle_{T}+\frac{1}{2}\langle-\hat{\mathbf{C}}_{i}\hat{\mathbf{C}}_{j}-\hat{\mathbf{C}}_{j}\hat{\mathbf{C}}_{i}\rangle_{T}+\frac{1}{2}\langle\hat{\mathbf{C}}_{j}^{2}\rangle_{T}}\right)_{\alpha\alpha';\beta'\beta}\nonumber \\
= & \sum_{\alpha'',\beta''}\left(e^{\frac{1}{2}\langle\hat{\mathbf{C}}_{i}^{2}\rangle_{T}}\right)_{\alpha\alpha''}\left(e^{\frac{1}{2}\langle-\hat{\mathbf{C}}_{i}\hat{\mathbf{C}}_{j}-\hat{\mathbf{C}}_{j}\hat{\mathbf{C}}_{i}\rangle_{T}}\right)_{\alpha''\alpha';\beta'\beta''}\left(e^{\frac{1}{2}\langle\hat{\mathbf{C}}_{j}^{2}\rangle_{T}}\right)_{\beta''\beta}.\label{eq:thermal_1}
\end{alignat}

The exponents above contains matrices $\langle\hat{\mathbf{C}}_{i}^{2}\rangle_{T},\:\langle\hat{\mathbf{C}}_{j}^{2}\rangle_{T},\;\langle\hat{\mathbf{C}}_{i}\hat{\mathbf{C}}_{j}\rangle_{T}$,
which do not commute with each other. Fortunately, by using the shorthand
notation $\left(\cdots\right)_{\alpha\alpha';\beta'\beta}$, those
matrix elements with index $\alpha\alpha';\beta'\beta$ do commute
between each other. They can be separated into independent exponents.
Thermal averages such as $\langle\hat{\mathbf{C}}_{i}^{2}\rangle_{T}$
can be calculated easily:
\begin{equation}
\langle\hat{\mathbf{C}}_{i}^{2}\rangle_{T}=\langle\sum_{\mathbf{q}}\sum_{\mathbf{q}'}\mathbf{\Lambda}_{i,\mathbf{q}}^{\dagger}\mathbf{\Lambda}_{i,\mathbf{q}'}\hat{\mathbf{b}}_{\mathbf{q}}^{\dagger}\hat{\mathbf{b}}_{\mathbf{q}'}+\sum_{\mathbf{q}}\sum_{\mathbf{q}'}\mathbf{\Lambda}_{i,\mathbf{q}'}\mathbf{\Lambda}_{i,\mathbf{q}}^{\dagger}\hat{\mathbf{b}}_{\mathbf{q}'}\hat{\mathbf{b}}_{\mathbf{q}}^{\dagger}\rangle_{T}=\sum_{\mathbf{q}}\left(\mathbf{\Lambda}_{i,\mathbf{q}}^{\dagger}\mathbf{\Lambda}_{i,\mathbf{q}}\right)\left(2N_{\mathbf{q}}+1\right)
\end{equation}
where $N_{\mathbf{q}}\equiv\left(\exp\left(\hbar\omega_{\mathbf{q}}/k_{\text{B}}T\right)-1\right)^{-1}$
is the thermally averaged phonon occupation number. The result of
$\langle\hat{\mathbf{C}}_{i}^{2}\rangle_{T}$ is guaranteed to be
a real number because of the symmetry relations $\mathbf{\Lambda}_{i,\mathbf{q}}=\mathbf{\Lambda}_{i,-\mathbf{q}}^{\dagger}$.
The thermal averages in Eq.~(\ref{eq:thermal_1}) can be calculated
as:
\begin{equation}
\langle\left(e^{\hat{\mathbf{C}}_{i}-\hat{\mathbf{C}}_{j}}\right)_{\alpha\alpha';\beta'\beta}\rangle_{T}\simeq\sum_{\alpha'',\beta''}\left(e^{-\sum_{\mathbf{q}}\left(N_{\mathbf{q}}+\frac{1}{2}\right)\cdot\left(\mathbf{\Lambda}_{i,\mathbf{q}}^{\dagger}\mathbf{\Lambda}_{i,\mathbf{q}}\right)}\right)_{\alpha\alpha''}\left(e^{\sum_{\mathbf{q}}\left(2N_{\mathbf{q}}+1\right)\left(\mathbf{\Lambda}_{i,\mathbf{q}}^{\dagger}\mathbf{\Lambda}_{j,\mathbf{q}}\right)}\right)_{\alpha''\alpha';\beta'\beta''}\left(e^{-\sum_{\mathbf{q}}\left(N_{\mathbf{q}}+\frac{1}{2}\right)\cdot\left(\mathbf{\Lambda}_{j,\mathbf{q}}^{\dagger}\mathbf{\Lambda}_{j,\mathbf{q}}\right)}\right)_{\beta''\beta}.
\end{equation}
The first and third term are easy to calculate. Then we expand the
middle exponential under the shorthand notation $\left(\cdots\right)_{\alpha''\alpha';\beta'\beta''}$:
\begin{alignat}{1}
 & \left(e^{\sum_{\mathbf{q}}\left(2N_{\mathbf{q}}+1\right)\left(\mathbf{\Lambda}_{i,\mathbf{q}}^{\dagger}\mathbf{\Lambda}_{j,\mathbf{q}}\right)}\right)_{\alpha''\alpha';\beta'\beta''}=\delta_{\alpha''\alpha'}\delta_{\beta'\beta''}+\frac{1}{1!}\sum_{\mathbf{q}}\left(2N_{\mathbf{q}}+1\right)\left(\mathbf{\Lambda}_{i,\mathbf{q}}^{\dagger}\right)_{\alpha''\alpha'}\left(\mathbf{\Lambda}_{j,\mathbf{q}}\right)_{\beta'\beta''}\nonumber \\
 & +\frac{1}{2!}\sum_{\mathbf{q}\mathbf{q}'}\left(2N_{\mathbf{q}}+1\right)\left(2N_{\mathbf{q}'}+1\right)\left(\mathbf{\Lambda}_{i,\mathbf{q}}^{\dagger}\mathbf{\Lambda}_{i,\mathbf{q}'}^{\dagger}\right)_{\alpha''\alpha'}\left(\mathbf{\Lambda}_{j,\mathbf{q}}\mathbf{\Lambda}_{j,\mathbf{q}'}\right)_{\beta'\beta''}+\cdots\nonumber \\
 & +\frac{1}{k!}\sum_{\mathbf{q}\mathbf{q}'\cdots\mathbf{q}^{\left(k-1\right)}}\underbrace{\left(2N_{\mathbf{q}}+1\right)\cdots\left(2N_{\mathbf{q}^{\left(k-1\right)}}+1\right)}_{k}\underset{k}{\underbrace{\left(\mathbf{\Lambda}_{i,\mathbf{q}}^{\dagger}\mathbf{\Lambda}_{i,\mathbf{q}'}^{\dagger}\cdots\mathbf{\Lambda}_{i,\mathbf{q}^{\left(k-1\right)}}^{\dagger}\right)_{\alpha''\alpha'}}}\underset{k}{\underbrace{\left(\mathbf{\Lambda}_{j,\mathbf{q}}\mathbf{\Lambda}_{j,\mathbf{q}'}\cdots\mathbf{\Lambda}_{j,\mathbf{q}^{\left(k-1\right)}}\right)_{\beta'\beta''}}}.
\end{alignat}
In each term, we combine elements with the same momentum $\mathbf{q}$:

\begin{alignat}{1}
 & \left(e^{\sum_{\mathbf{q}}\left(2N_{\mathbf{q}}+1\right)\left(\mathbf{\Lambda}_{i,\mathbf{q}}^{\dagger}\mathbf{\Lambda}_{j,\mathbf{q}}\right)}\right)_{\alpha''\alpha';\beta'\beta''}=\delta_{\alpha''\alpha'}\delta_{\beta'\beta''}+\frac{1}{1!}\sum_{\mathbf{q}}\left(2N_{\mathbf{q}}+1\right)\left(\mathbf{\Lambda}_{i,\mathbf{q}}^{\dagger}\right)_{\alpha''\alpha'}\left(\mathbf{\Lambda}_{j,\mathbf{q}}\right)_{\beta'\beta''}\nonumber \\
 & +\frac{1}{2!}\sum_{\alpha^{\left(3\right)},\beta^{\left(3\right)}}\left(\sum_{\mathbf{q}}\left(2N_{\mathbf{q}}+1\right)\left(\mathbf{\Lambda}_{i,\mathbf{q}}^{\dagger}\right)_{\alpha''\alpha^{\left(3\right)}}\left(\mathbf{\Lambda}_{j,\mathbf{q}}\right)_{\beta'\beta^{\left(3\right)}}\right)\left(\sum_{\mathbf{q}'}\left(2N_{\mathbf{q}'}+1\right)\left(\mathbf{\Lambda}_{i,\mathbf{\mathbf{q}'}}^{\dagger}\right)_{\alpha^{\left(3\right)}\alpha'}\left(\mathbf{\Lambda}_{j,\mathbf{\mathbf{q}'}}\right)_{\beta^{\left(3\right)}\beta''}\right)+\cdots\nonumber \\
 & +\frac{1}{k!}\underbrace{\sum_{\alpha^{\left(3\right)},\beta^{\left(3\right)}}\cdots\sum_{\alpha^{\left(k+1\right)},\beta^{\left(k+1\right)}}}_{k-1}\nonumber \\
 & \quad\cdot\underbrace{\left(\sum_{\mathbf{q}}\left(2N_{\mathbf{q}}+1\right)\left(\mathbf{\Lambda}_{i,\mathbf{q}}^{\dagger}\right)_{\alpha''\alpha^{\left(3\right)}}\left(\mathbf{\Lambda}_{j,\mathbf{q}}\right)_{\beta'\beta^{\left(3\right)}}\right)\cdots\left(\sum_{\mathbf{q}^{\left(k-1\right)}}\left(2N_{\mathbf{q}^{\left(k-1\right)}}+1\right)\left(\mathbf{\Lambda}_{i,\mathbf{\mathbf{q}^{\left(k-1\right)}}}^{\dagger}\right)_{\alpha^{\left(k+1\right)}\alpha'}\left(\mathbf{\Lambda}_{j,\mathbf{\mathbf{q}^{\left(k-1\right)}}}\right)_{\beta^{\left(k+1\right)}\beta''}\right)}_{k}\label{eq:expG}
\end{alignat}
In order to simplify last part in Eq.~(\ref{eq:expG}), we define
4th rank tensors \textbf{$\mathbf{G}_{ij}$} with elements:
\begin{equation}
\left(\mathbf{G}_{ij}\right)_{\alpha''\alpha'\beta'\beta''}\equiv\sum_{\mathbf{q}}\left(2N_{\mathbf{q}}+1\right)\left(\mathbf{\Lambda}_{i,\mathbf{q}}^{\dagger}\right)_{\alpha''\alpha'}\left(\mathbf{\Lambda}_{j,\mathbf{q}}\right)_{\beta'\beta''},\label{eq:matrices_G}
\end{equation}
and
\begin{equation}
\left(\mathbf{G}_{ij}^{2}\right)_{\alpha\alpha'\beta\beta'}\equiv\sum_{\alpha^{\prime\prime}\beta^{\prime\prime}}\left(\mathbf{G}_{ij}\right)_{\alpha\alpha''\beta\beta''}\left(\mathbf{G}_{ij}\right)_{\alpha''\alpha'\beta''\beta'}.
\end{equation}
Then the eq.~(\ref{eq:expG}) can be written as:
\begin{alignat}{1}
 & \left(e^{\sum_{\mathbf{q}}\left(2N_{\mathbf{q}}+1\right)\left(\mathbf{\Lambda}_{i,\mathbf{q}}^{\dagger}\mathbf{\Lambda}_{j,\mathbf{q}}\right)}\right)_{\alpha''\alpha';\beta'\beta''}\nonumber \\
= & \:\delta_{\alpha''\alpha'}\delta_{\beta'\beta''}+\frac{1}{1!}\left(\mathbf{G}_{ij}\right)_{\alpha''\alpha'\beta'\beta''}+\frac{1}{2!}\left(\mathbf{G}_{ij}^{2}\right)_{\alpha''\alpha'\beta'\beta''}+\frac{1}{k!}\left(\mathbf{G}_{ij}^{k}\right)_{\alpha''\alpha'\beta'\beta''}\nonumber \\
= & \left(e^{\mathbf{G}_{ij}}\right)_{\alpha''\alpha'\beta'\beta''},
\end{alignat}
where $\left(e^{\mathbf{G}_{ij}}\right)$ are also 4th rank tensors.

The tensors \textbf{$\mathbf{G}_{ij}$} involve only numbers, and
thus $\left(e^{\mathbf{G}_{ij}}\right)$ can be calculated exactly.
Finally, the first thermal average term can be calculated as:
\begin{alignat}{1}
\langle\left(\hat{\mathbf{X}}_{i}^{\dagger}\mathcal{F}\hat{\mathbf{X}}_{j}\right)_{\alpha\beta}\rangle_{T}= & \sum_{\alpha'\beta'}\left(\mathcal{F}\right)_{\alpha'\beta'}\langle\left(e^{\hat{\mathbf{C}}_{i}-\hat{\mathbf{C}}_{j}}\right)_{\alpha\alpha';\beta'\beta}\rangle_{T}\nonumber \\
\langle\left(e^{\hat{\mathbf{C}}_{i}-\hat{\mathbf{C}}_{j}}\right)_{\alpha\alpha';\beta'\beta}\rangle_{T}= & \sum_{\alpha'',\beta''}\left(e^{-\sum_{\mathbf{q}}\left(N_{\mathbf{q}}+\frac{1}{2}\right)\cdot\left(\mathbf{\Lambda}_{i,\mathbf{q}}^{\dagger}\mathbf{\Lambda}_{i,\mathbf{q}}\right)}\right)_{\alpha\alpha''}\left(e^{\mathbf{G}_{ij}}\right)_{\alpha''\alpha'\beta'\beta''}\left(e^{-\sum_{\mathbf{q}}\left(N_{\mathbf{q}}+\frac{1}{2}\right)\cdot\left(\mathbf{\Lambda}_{j,\mathbf{q}}^{\dagger}\mathbf{\Lambda}_{j,\mathbf{q}}\right)}\right)_{\beta''\beta}.\label{eq:thermal_1_end}
\end{alignat}

\subsubsection*{Part II: \textup{$\langle\left(\hat{\mathbf{X}}_{i}^{\dagger}\hat{\mathbf{b}}_{\mathbf{q}}^{\dagger}\mathbf{M}_{i,\mathbf{q}}^{\dagger}\hat{\mathbf{X}}_{i}\right)\rangle_{T}$} }

Here we calculate the second type of thermal averages by the relation
$f\left(x=1\right)=\int_{0}^{1}f'\left(x\right)dx+f\left(x=0\right)$:
\begin{alignat}{1}
 & \langle\left(\hat{\mathbf{X}}_{i}^{\dagger}\hat{\mathbf{b}}_{\mathbf{q}}^{\dagger}\mathbf{M}_{i,\mathbf{q}}^{\dagger}\hat{\mathbf{X}}_{i}\right)\rangle_{T}=\langle\left(e^{\hat{\mathbf{C}}_{i}}\hat{\mathbf{b}}_{\mathbf{q}}^{\dagger}e^{-\hat{\mathbf{C}}_{i}}\right)\left(e^{\hat{\mathbf{C}}_{i}}\mathbf{M}_{i,\mathbf{q}}^{\dagger}e^{-\hat{\mathbf{C}}_{i}}\right)\rangle_{T}\nonumber \\
= & \langle\int_{0}^{1}dx\frac{d}{dx}\left(e^{x\hat{\mathbf{C}}_{i}}\hat{\mathbf{b}}_{\mathbf{q}}^{\dagger}e^{-x\hat{\mathbf{C}}_{i}}e^{\hat{\mathbf{C}}_{i}}\mathbf{M}_{i,\mathbf{q}}^{\dagger}e^{-\hat{\mathbf{C}}_{i}}\right)\rangle_{T}+\langle\hat{\mathbf{b}}_{\mathbf{q}}^{\dagger}e^{\hat{\mathbf{C}}_{i}}\mathbf{M}_{i,\mathbf{q}}^{\dagger}e^{-\hat{\mathbf{C}}_{i}}\rangle_{T}\nonumber \\
= & -\int_{0}^{1}dx\langle e^{x\hat{\mathbf{C}}_{i}}\mathbf{\Lambda}_{i,\mathbf{q}}e^{-x\hat{\mathbf{C}}_{i}}e^{\hat{\mathbf{C}}_{i}}\mathbf{M}_{i,\mathbf{q}}^{\dagger}e^{-\hat{\mathbf{C}}_{i}}\rangle_{T}+\langle\hat{\mathbf{b}}_{\mathbf{q}}^{\dagger}e^{\hat{\mathbf{C}}_{i}}\mathbf{M}_{i,\mathbf{q}}^{\dagger}e^{-\hat{\mathbf{C}}_{i}}\rangle_{T}\label{eq:thermal_2}
\end{alignat}
where we have used the relation $\left[\hat{\mathbf{C}}_{i},\;\hat{\mathbf{b}}_{\mathbf{q}}^{\dagger}\right]=-\mathbf{\Lambda}_{i,\mathbf{q}}$.
The first term can be calculated by separating out the bath terms:
\begin{alignat}{1}
 & \langle\left(e^{x\hat{\mathbf{C}}_{i}}\mathbf{\Lambda}_{i,\mathbf{q}}e^{-x\hat{\mathbf{C}}_{i}}e^{\hat{\mathbf{C}}_{i}}\mathbf{M}_{i,\mathbf{q}}^{\dagger}e^{-\hat{\mathbf{C}}_{i}}\right)_{\alpha\beta}\rangle_{T}\nonumber \\
= & \sum_{\alpha'\gamma\gamma'\beta'}\langle\left(e^{x\hat{\mathbf{C}}_{i}}\right)_{\alpha\alpha'}\left(\mathbf{\Lambda}_{i,\mathbf{q}}\right)_{\alpha'\gamma}\left(e^{\left(1-x\right)\hat{\mathbf{C}}_{i}}\right)_{\gamma\gamma'}\left(\mathbf{M}_{i,\mathbf{q}}^{\dagger}\right)_{\gamma'\beta'}\left(e^{-\hat{\mathbf{C}}_{i}}\right)_{\beta'\beta}\rangle_{T}\nonumber \\
= & \sum_{\alpha'\gamma\gamma'\beta'}\left(\mathbf{\Lambda}_{i,\mathbf{q}}\right)_{\alpha'\gamma}\left(\mathbf{M}_{i,\mathbf{q}}^{\dagger}\right)_{\gamma'\beta'}\cdot\langle\left(e^{x\hat{\mathbf{C}}_{i}}\right)_{\alpha\alpha'}\left(e^{\left(1-x\right)\hat{\mathbf{C}}_{i}}\right)_{\gamma\gamma'}\left(e^{-\hat{\mathbf{C}}_{i}}\right)_{\beta'\beta}\rangle_{T}.
\end{alignat}
This element can be calculated by the same method as we applied for
$\langle\left(e^{\hat{\mathbf{C}}_{i}}\right)_{\alpha\alpha'}\left(e^{-\hat{\mathbf{C}}_{j}}\right)_{\beta'\beta}\rangle_{T}$
in the previous section. Under the same shorthand notation:
\begin{alignat}{1}
 & \langle\left(e^{x\hat{\mathbf{C}}_{i}}\right)_{\alpha\alpha'}\left(e^{\left(1-x\right)\hat{\mathbf{C}}_{j}}\right)_{\gamma\gamma'}\left(e^{-\hat{\mathbf{C}}_{k}}\right)_{\beta'\beta}\rangle_{T}=\langle\left(e^{x\hat{\mathbf{C}}_{i}+\left(1-x\right)\hat{\mathbf{C}}_{j}-\hat{\mathbf{C}}_{k}}\right)_{\alpha\alpha';\gamma\gamma';\beta'\beta}\rangle_{T}\nonumber \\
\simeq & \left(e^{\frac{1}{2}\langle\left(x\hat{\mathbf{C}}_{i}+\left(1-x\right)\hat{\mathbf{C}}_{j}-\hat{\mathbf{C}}_{k}\right)^{2}\rangle_{T}}\right)_{\alpha\alpha';\gamma\gamma';\beta'\beta}\nonumber \\
= & \sum_{\alpha^{2}\alpha^{3}}\sum_{\gamma^{2}\gamma^{3}}\sum_{\beta^{2}\beta^{3}}\left(e^{\frac{1}{2}\langle x^{2}\hat{\mathbf{C}}_{i}^{2}\rangle_{T}}\right)_{\alpha\alpha^{\left(2\right)}}\left(e^{\frac{1}{2}\langle\left(1-x\right)^{2}\hat{\mathbf{C}}_{j}^{2}\rangle_{T}}\right)_{\gamma\gamma^{\left(2\right)}}\left(e^{\frac{1}{2}\langle\hat{\mathbf{C}}_{k}^{2}\rangle_{T}}\right)_{\beta'\beta^{\left(2\right)}}\nonumber \\
 & \cdot\left(e^{\frac{1}{2}\langle x\hat{\mathbf{C}}_{i}\left(1-x\right)\hat{\mathbf{C}}_{j}\rangle_{T}}\right)_{\alpha^{\left(2\right)}\alpha^{\left(3\right)};\gamma^{\left(2\right)}\gamma^{\left(3\right)}}\left(e^{-\frac{1}{2}\langle x\hat{\mathbf{C}}_{i}\hat{\mathbf{C}}_{k}\rangle_{T}}\right)_{\alpha^{\left(3\right)}\alpha';\beta^{\left(2\right)}\beta^{\left(3\right)}}\left(e^{-\frac{1}{2}\langle\left(1-x\right)\hat{\mathbf{C}}_{j}\hat{\mathbf{C}}_{k}\rangle_{T}}\right)_{\gamma^{\left(3\right)}\gamma';\beta^{\left(3\right)}\beta}\label{eq:thermal_2_1}
\end{alignat}
with $\alpha,\beta,\gamma$ applied separately for $x\hat{\mathbf{C}}_{i}$,
$\left(1-x\right)\hat{\mathbf{C}}_{j}$ and $-\hat{\mathbf{C}}_{k}$.
We have introduced the different indices $i$, $j$ and $k$ so that
we can keep track of the terms required in applying the short-hand
notation. In the end, we will set $i=j=k$ to recover the desired
result. Using the results from previous section, these thermal averages
will be:
\begin{alignat}{1}
 & \langle\left(e^{x\hat{\mathbf{C}}_{i}}\right)_{\alpha\alpha'}\left(e^{\left(1-x\right)\hat{\mathbf{C}}_{i}}\right)_{\gamma\gamma'}\left(e^{-\hat{\mathbf{C}}_{i}}\right)_{\beta'\beta}\rangle_{T}\nonumber \\
= & \sum_{\alpha^{2}\alpha^{3}}\sum_{\gamma^{2}\gamma^{3}}\sum_{\beta^{2}\beta^{3}}\left(e^{-\sum_{\mathbf{q}}\left(N_{\mathbf{q}}+\frac{1}{2}\right)\cdot x^{2}\cdot\mathbf{\Lambda}_{\mathbf{q}}^{\dagger}\mathbf{\Lambda}_{\mathbf{q}}}\right)_{\alpha\alpha^{2}}\left(e^{-\sum_{\mathbf{q}}\left(N_{\mathbf{q}}+\frac{1}{2}\right)\cdot\left(1-x\right)^{2}\cdot\mathbf{\Lambda}_{\mathbf{q}}^{\dagger}\mathbf{\Lambda}_{\mathbf{q}}}\right)_{\gamma\gamma^{2}}\left(e^{-\sum_{\mathbf{q}}\left(N_{\mathbf{q}}+\frac{1}{2}\right)\cdot\mathbf{\Lambda}_{\mathbf{q}}^{\dagger}\mathbf{\Lambda}_{\mathbf{q}}}\right)_{\beta'\beta^{2}}\nonumber \\
 & \cdot\left(e^{-x\left(1-x\right)\mathbf{G}_{ij}}\right)_{\alpha^{\left(2\right)}\alpha^{\left(3\right)}\gamma\gamma^{\left(3\right)}}\left(e^{x\mathbf{G}_{ik}}\right)_{\alpha^{\left(3\right)}\alpha'\beta^{\left(2\right)}\beta^{\left(3\right)}}\left(e^{\left(1-x\right)\mathbf{G}_{jk}}\right)_{\gamma^{\left(3\right)}\gamma'\beta^{\left(3\right)}\beta}\label{eq:thermal_2_2}
\end{alignat}
with $\mathbf{G}_{ij}$ from Eq.~(\ref{eq:matrices_G}).

On the other hand, the second term $\langle\hat{\mathbf{b}}_{\mathbf{q}}^{\dagger}e^{\hat{\mathbf{C}}_{i}}\mathbf{M}_{i,\mathbf{q}}^{\dagger}e^{-\hat{\mathbf{C}}_{i}}\rangle_{T}$
in Eq.~(\ref{eq:thermal_2}) can be calculated as:
\begin{equation}
\langle\left(\hat{\mathbf{b}}_{\mathbf{q}}^{\dagger}e^{\hat{\mathbf{C}}_{i}}\mathbf{M}_{i,\mathbf{q}}^{\dagger}e^{-\hat{\mathbf{C}}_{i}}\right)_{\alpha\beta}\rangle_{T}=\sum_{\alpha'\beta'}\left(\mathbf{M}_{i,\mathbf{q}}^{\dagger}\right)_{\alpha'\beta'}\langle\hat{b}_{\mathbf{q}}^{\dagger}\left(e^{\hat{\mathbf{C}}_{i}-\hat{\mathbf{C}}_{i}}\right)_{\alpha\alpha';\beta'\beta}\rangle_{T}.
\end{equation}
From the results in \cite{Munn1985b}, the elements such as $\langle\hat{\mathbf{b}}_{\mathbf{q}}^{\dagger}\left(e^{\hat{\mathbf{C}}_{i}-\hat{\mathbf{C}}_{i}}\right)\rangle_{T}$
can be obtained by differentiating $\left(e^{\hat{\mathbf{C}}_{i}-\hat{\mathbf{C}}_{i}}\right)$
before taking the thermal average. On the other hand, since the thermal
average only involves the phonon degrees of freedom, we can switch
the order and differentiate $\langle\left(e^{\hat{\mathbf{C}}_{i}-\hat{\mathbf{C}}_{i}}\right)\rangle_{T}$
after taking the thermal average. By comparing the two steps, we obtain
the relation:
\begin{equation}
\langle\hat{\mathbf{b}}_{\mathbf{q}}^{\dagger}\left(e^{\hat{\mathbf{C}}_{i}-\hat{\mathbf{C}}_{i}}\right)\rangle_{T}=-\langle\hat{\mathbf{b}}_{\mathbf{q}}^{\dagger}\sum_{\mathbf{\mathbf{q}'}}\mathbf{\Lambda}_{i,\mathbf{\mathbf{q}'}}\hat{\mathbf{b}}_{\mathbf{\mathbf{q}'}}\rangle_{T}\langle\left(e^{\hat{\mathbf{C}}_{i}-\hat{\mathbf{C}}_{i}}\right)\rangle_{T}+\langle\left(e^{\hat{\mathbf{C}}_{i}-\hat{\mathbf{C}}_{i}}\right)\rangle_{T}\langle\hat{\mathbf{b}}_{\mathbf{\mathbf{q}}}^{\dagger}\sum_{\mathbf{\mathbf{q}'}}\mathbf{\Lambda}_{i,\mathbf{\mathbf{q}'}}\hat{\mathbf{b}}_{\mathbf{\mathbf{q}'}}\rangle_{T},
\end{equation}
which can be calculated by the results in Eq.~(\ref{eq:thermal_1_end}).
As an example, we show a simplified calculation in Eq.(\ref{eq:<b_K>}),
where the transformation matrices $\mathbf{\Lambda}_{i,\mathbf{q}}$
are diagonal. 

We finally obtain the second term in Eq.~(\ref{eq:thermal_2}): 
\begin{alignat}{1}
 & \langle\hat{\mathbf{b}}_{\mathbf{q}}^{\dagger}e^{\hat{\mathbf{C}}_{i}}\mathbf{M}_{i,\mathbf{q}}^{\dagger}e^{-\hat{\mathbf{C}}_{i}}\rangle_{T}\nonumber \\
= & \sum_{\alpha'\beta'}N_{\mathbf{q}}\left(\mathbf{M}_{i,\mathbf{q}}^{\dagger}\right)_{\alpha'\beta'}\left[\sum_{\alpha''}\left(-\mathbf{\Lambda}_{i,\mathbf{q}}\right)_{\alpha\alpha''}\langle\left(e^{\hat{\mathbf{C}}_{i}-\hat{\mathbf{C}}_{i}}\right)_{\alpha''\alpha';\beta'\beta}\rangle_{T}+\sum_{\beta''}\langle\left(e^{\hat{\mathbf{C}}_{i}-\hat{\mathbf{C}}_{i}}\right)_{\alpha\alpha';\beta'\beta''}\rangle_{T}\left(\mathbf{\Lambda}_{i,\mathbf{q}}\right)_{\beta''\beta}\right].\label{eq:thermal_2_3}
\end{alignat}

\subsubsection*{Part III: \textup{$\langle\left(\hat{\mathbf{X}}_{i}^{\dagger}\hat{\mathbf{b}}_{\mathbf{q}}^{\dagger}\hat{\mathbf{b}}_{\mathbf{q}}\hat{\mathbf{X}}_{i}\right)\rangle_{T}$}}

This term can be calculated by a similar integral method as in the
previous section:
\begin{alignat}{1}
 & \langle\left(\hat{\mathbf{X}}_{i}^{\dagger}\hat{\mathbf{b}}_{\mathbf{q}}^{\dagger}\hat{\mathbf{b}}_{\mathbf{q}}\hat{\mathbf{X}}_{i}\right)\rangle_{T}=\int_{0}^{1}dy\langle\frac{d}{dy}\left(e^{y\hat{\mathbf{C}}_{i}}\hat{\mathbf{b}}_{\mathbf{q}}^{\dagger}\hat{\mathbf{b}}_{\mathbf{q}}e^{-y\hat{\mathbf{C}}_{i}}\right)\rangle_{T}+\langle\hat{\mathbf{b}}_{\mathbf{q}}^{\dagger}\hat{\mathbf{b}}_{\mathbf{q}}\rangle_{T}\nonumber \\
= & -\int_{0}^{1}dy\langle e^{y\hat{\mathbf{C}}_{i}}\hat{\mathbf{b}}_{\mathbf{q}}^{\dagger}\mathbf{\Lambda}_{i,\mathbf{q}}^{\dagger}e^{-y\hat{\mathbf{C}}_{i}}+h.c.\rangle_{T}+\langle\hat{\mathbf{b}}_{\mathbf{q}}^{\dagger}\hat{\mathbf{b}}_{\mathbf{q}}\rangle_{T}\label{eq:thermal_3}
\end{alignat}
Here the thermal average $\langle e^{y\hat{\mathbf{C}}_{i}}\hat{\mathbf{b}}_{\mathbf{q}}^{\dagger}\mathbf{\Lambda}_{i,\mathbf{q}}^{\dagger}e^{-y\hat{\mathbf{C}}_{i}}\rangle_{T}$
can be obtained by changing $\hat{\mathbf{C}}_{i}$ to $y\hat{\mathbf{C}}_{i}$
and $\mathbf{M}_{i,\mathbf{q}}$ to $\mathbf{\Lambda}_{i,\mathbf{q}}$
in Eq.~(\ref{eq:thermal_2}):
\begin{alignat}{1}
 & \langle\left(e^{y\hat{\mathbf{C}}_{i}}\hat{\mathbf{b}}_{\mathbf{q}}^{\dagger}\mathbf{\Lambda}_{i,\mathbf{q}}^{\dagger}e^{-y\hat{\mathbf{C}}_{i}}\right)_{\alpha\beta}\rangle_{T}\nonumber \\
= & -\int_{0}^{1}dx\left[\sum_{\alpha'\gamma\gamma'\beta'}y\left(\mathbf{\Lambda}_{i,\mathbf{q}}\right)_{\alpha'\gamma}\left(\mathbf{\Lambda}_{i,\mathbf{q}}^{\dagger}\right)_{\gamma'\beta'}\cdot\langle\left(e^{xy\hat{\mathbf{C}}_{i}}\right)_{\alpha\alpha'}\left(e^{\left(1-x\right)y\hat{\mathbf{C}}_{i}}\right)_{\gamma\gamma'}\left(e^{-y\hat{\mathbf{C}}_{i}}\right)_{\beta'\beta}\rangle_{T}\right]\nonumber \\
 & +N_{\mathbf{q}}\sum_{\alpha'\beta'}\left(\mathbf{\Lambda}_{i,\mathbf{q}}^{\dagger}\right)_{\alpha'\beta'}y\cdot\left[\sum_{\alpha''}\left(-\mathbf{\Lambda}_{i,\mathbf{q}}\right)_{\alpha\alpha''}\langle\left(e^{y\hat{\mathbf{C}}_{i}-y\hat{\mathbf{C}}_{i}}\right)_{\alpha''\alpha';\beta'\beta}\rangle_{T}+\sum_{\beta''}\langle\left(e^{y\hat{\mathbf{C}}_{i}-y\hat{\mathbf{C}}_{i}}\right)_{\alpha\alpha';\beta'\beta''}\rangle_{T}\left(\mathbf{\Lambda}_{i,\mathbf{q}}\right)_{\beta''\beta}\right].
\end{alignat}

\subsubsection*{Part IV: \textup{$\langle\left(\hat{\mathbf{X}}_{i}^{\dagger}\hat{\mathbf{b}}_{\mathbf{q}}^{\dagger}\hat{\mathbf{X}}_{i}^{\dagger}-\hat{\mathbf{b}}_{\mathbf{q}}^{\dagger}\right)_{\alpha\beta}\left(\hat{\mathbf{X}}_{i}^{\dagger}\hat{\mathbf{b}}_{\mathbf{q}}\hat{\mathbf{X}}_{i}^{\dagger}-\hat{\mathbf{b}}_{\mathbf{q}}\right)_{\alpha'\beta'}\rangle_{T}$}}

This last term describes bath-induced interactions between polarons.
By using a relation similar to Eq.~(\ref{eq:thermal_2}) and applying
it only to the $\hat{\mathbf{X}}_{i}^{\dagger}\hat{\mathbf{b}}_{\mathbf{q}}^{\dagger}\hat{\mathbf{X}}_{i}^{\dagger}$
term, we obtain
\begin{equation}
\left(\hat{\mathbf{X}}_{i}^{\dagger}\hat{\mathbf{b}}_{\mathbf{q}}^{\dagger}\hat{\mathbf{X}}_{i}^{\dagger}-\hat{\mathbf{b}}_{\mathbf{q}}^{\dagger}\right)=\int_{0}^{1}dx\frac{d}{dx}\left(e^{x\hat{\mathbf{C}}_{i}}\hat{\mathbf{b}}_{\mathbf{q}}^{\dagger}e^{-x\hat{\mathbf{C}}_{i}}\right)=-\int_{0}^{1}dx\left(e^{x\hat{\mathbf{C}}_{i}}\mathbf{\Lambda}_{i,\mathbf{q}}e^{-x\hat{\mathbf{C}}_{i}}\right),
\end{equation}
and
\begin{alignat}{1}
 & \langle\left(\hat{\mathbf{X}}_{i}^{\dagger}\hat{\mathbf{b}}_{\mathbf{q}}^{\dagger}\hat{\mathbf{X}}_{i}^{\dagger}-\hat{\mathbf{b}}_{\mathbf{q}}^{\dagger}\right)_{\alpha\beta}\left(\hat{\mathbf{X}}_{i}^{\dagger}\hat{\mathbf{b}}_{\mathbf{q}}\hat{\mathbf{X}}_{i}^{\dagger}-\hat{\mathbf{b}}_{\mathbf{q}}\right)_{\alpha'\beta'}\rangle_{T}=\int_{0}^{1}\int_{0}^{1}dxdy\langle\left(e^{x\hat{\mathbf{C}}_{i}}\mathbf{\Lambda}_{i,\mathbf{q}}e^{-x\hat{\mathbf{C}}_{i}}\right)_{\alpha\beta}\left(e^{y\hat{\mathbf{C}}_{i}}\mathbf{\Lambda}_{i,\mathbf{q}}^{\dagger}e^{-y\hat{\mathbf{C}}_{i}}\right)_{\alpha'\beta'}\rangle_{T}\nonumber \\
= & \int_{0}^{1}\int_{0}^{1}dxdy\sum_{\gamma\gamma'}\sum_{\delta\delta'}\left(\mathbf{\Lambda}_{i,\mathbf{q}}\right)_{\gamma\gamma'}\left(\mathbf{\Lambda}_{i,\mathbf{\mathbf{q}}}^{\dagger}\right)_{\delta\delta'}\cdot\langle\left(e^{x\hat{\mathbf{C}}_{i}}\right)_{\alpha\gamma}\left(e^{-x\hat{\mathbf{C}}_{i}}\right)_{\gamma'\beta}\left(e^{y\hat{\mathbf{C}}_{i}}\right)_{\alpha'\delta}\left(e^{-y\hat{\mathbf{C}}_{i}}\right)_{\delta'\beta'}\rangle_{T}\label{eq:thermal_4}
\end{alignat}
This can thus be calculated by the same method as in Eqs. (\ref{eq:thermal_2_1},
\ref{eq:thermal_2_2}).

In addition, we would like to point out that the thermal average calculation
in Eq.~(\ref{eq:thermal_1}) is exact only when the matrices $\mathbf{\mathbf{\Lambda}}_{\mathbf{q}}$
commute with each other: $\left[\mathbf{\Lambda}_{\mathbf{q}},\;\mathbf{\Lambda}_{\mathbf{\mathbf{q}'}}\right]=0$,
even though the other derivations are exact for an arbitrary choice
of $\mathbf{\mathbf{\Lambda}}_{\mathbf{q}}$. In order to maintain
good control over our two-band polaron transformation method, we can
also limit the variational parameters $\mathbf{\mathbf{\Lambda}}_{\mathbf{q}}$
to be a set of \textit{commuting matrices}. Though the ground state
energy and the contribution from the incoherent part might thus increase,
this transformation has the great advantage of an exactly solvable
coherent part. Another benefit of using commuting matrices is that
the transformed creation/annihilation operators take a simplified
form $e^{\hat{S}}\hat{b}_{\mathbf{q}}e^{-\hat{S}}=\hat{b}_{\mathbf{q}}-\sum_{i,\alpha,\beta}\left(\Lambda_{i,\mathbf{q}}^{\alpha\beta}\right)^{*}\hat{a}_{i}^{\alpha\dagger}\hat{a}_{i}^{\beta}$.

\section{Lindblad equation\label{sub:App_Lindblad_polaron}}

After the polaron transformation, the Hamiltonian can be separated
into a coherent part Eq.~(\ref{eq:coherent}) and an incoherent part
Eq.~(\ref{eq:incoherent}). This Hamiltonian describes an quantum
system (coherent part) coupled to a bosonic bath via interactions
(incoherent part). We apply the Lindblad master equation to include
the incoherent part.

In the interaction picture, due to $J^{0};\:J^{1}\ll\varepsilon_{{\rm P}}^{\Delta}$
for a deep lattice, the polaron operators can be approximated as 
\[
\hat{a}_{i}^{\alpha}\left(t\right)\approx\hat{a}_{i}^{\alpha}e^{-i\varepsilon_{{\rm P}}^{\alpha}t/\hbar};\quad\hat{a}_{i}^{\alpha\dagger}\left(t\right)\approx\hat{a}_{i}^{\alpha\dagger}e^{+i\varepsilon_{{\rm P}}^{\alpha}t/\hbar},
\]
where $\varepsilon_{{\rm P}}^{\alpha}$ is the renormalized polaron
on-site energy. The interaction (incoherent part) in Eq.~(\ref{eq:incoherent})
is given by: 
\begin{alignat}{1}
H_{\text{inc}}\left(t\right)\approx & \sum_{\alpha}\sum_{i,j}\left\{ B_{i,j}^{\alpha\alpha}\left(t\right)\right\} \hat{a}_{i}^{\alpha\dagger}\hat{a}_{j}^{\alpha}+\sum_{\alpha\neq\beta}\sum_{i}\left\{ B_{i,i}^{\alpha\beta}\left(t\right)\right\} \hat{a}_{i}^{\alpha\dagger}\hat{a}_{i}^{\beta}\cdot e^{-i\left(\beta-\alpha\right)\varepsilon_{{\rm P}}^{\Delta}t/\hbar};\nonumber \\
B_{i,j}^{\alpha\alpha}\left(t\right)\equiv & -\delta_{j,i\pm1}J_{{\rm P}}^{\alpha}\hat{T}_{i,j}^{\alpha\alpha}\left(t\right)+\delta_{i,j}\sum_{\mathbf{q}}\hbar\omega_{\mathbf{q}}\left(\hat{b}_{\mathbf{q}}^{\dagger}\left(t\right)+\hat{b}_{-\mathbf{q}}\left(t\right)\right)\left(\mathbf{M}_{i,\mathbf{q}}^{\dagger}-\mathbf{\Lambda}_{i,\mathbf{q}}^{\dagger}\right)_{\alpha\alpha};\\
B_{i,i}^{\alpha\beta}\left(t\right)\equiv & \sum_{\mathbf{q}}\hbar\omega_{\mathbf{q}}\left[\hat{b}_{\mathbf{q}}^{\dagger}\left(t\right)\left(\mathbf{M}_{i,\mathbf{q}}^{\dagger}\right)_{\alpha\beta}\left(\hat{K}_{i,i}^{\alpha\beta}\left(t\right)\right)+\left(\mathbf{M}_{i,\mathbf{q}}\right)_{\beta\alpha}\left(\hat{K}_{i,i}^{\beta\alpha}\left(t\right)\right)^{*}\hat{b}_{\mathbf{q}}\left(t\right)\right]\nonumber \\
+ & \sum_{\mathbf{q}}\hbar\omega_{\mathbf{q}}\left[\left(\Lambda_{\mathbf{q}}^{\alpha}\right)\left(\mathbf{M}_{\mathbf{q}}^{\dagger}\right)_{\alpha\beta}\left(\hat{T}_{i,i}^{\alpha\beta}\left(t\right)\right)+\left(\Lambda_{\mathbf{q}}^{\beta}\right)^{*}\left(\mathbf{M}_{\mathbf{q}}\right)_{\beta\alpha}\left(\hat{T}_{i,i}^{\beta\alpha}\left(t\right)\right)^{*}\right],
\end{alignat}
where $\hat{K}_{i,i}^{\alpha\beta}\left(t\right)$ and $\hat{T}_{i,i}^{\alpha\beta}\left(t\right)$
are also in the interaction picture. The first term in $H_{\text{inc}}\left(t\right)$
describes intra-band dynamics, while the second term describes inter-band
dynamics. In both processes, the polaron will emit (absorb) phonons
to (from) the bath. Using the general form of the Lindblad equation
in Eq.~(\ref{eq:Lindblad}), we can define the decoherence factor
$\gamma$ as:
\begin{alignat}{1}
\gamma_{ij;i'j'}^{\alpha\beta;\alpha'\beta'}\left(\omega\right)= & \int_{-\infty}^{\infty}d\tau e^{i\omega\tau}\langle B_{ij}^{\alpha\beta\dagger}\left(\tau\right)B_{i'j'}^{\alpha'\beta'}\left(0\right)\rangle_{T}\nonumber \\
= & 2\cdot{\bf \text{Re}}\left(\Gamma_{ij;i'j'}^{\alpha\beta;\alpha'\beta'}\left(\omega\right)\right)\equiv2\cdot{\bf \text{Re}}\int_{0}^{\infty}d\tau e^{i\omega\tau}g_{ij;i'j'}^{\alpha\beta;\alpha'\beta'}\left(\tau\right).
\end{alignat}
By using the rotating wave approximation (RWA), we decouple intra-
and inter-band dynamics in the Lindblad equation:
\begin{alignat}{1}
\mathcal{L}_{I}\left[\rho\left(t\right)\right]= & \sum_{\alpha,\alpha'}\sum_{ij}\sum_{i'j'}\gamma_{ij;i'j'}^{\alpha\alpha;\alpha'\alpha'}\left(C_{i'j'}^{\alpha'\alpha'}\rho C_{ij}^{\alpha\alpha\dagger}-\frac{1}{2}\left\{ C_{ij}^{\alpha\alpha\dagger}C_{i'j'}^{\alpha'\alpha'},\;\rho\right\} \right)\nonumber \\
+ & \sum_{i}\sum_{i'}\gamma_{ii;i'i'}^{01;01}\left(C_{i'i'}^{01}\rho C_{ii}^{01\dagger}-\frac{1}{2}\left\{ C_{ii}^{01\dagger}C_{i'i'}^{01},\;\rho\right\} \right)\nonumber \\
+ & \sum_{i}\sum_{i'}\gamma_{ii;i'i'}^{10;10}\left(C_{i'i'}^{10}\rho C_{ii}^{10\dagger}-\frac{1}{2}\left\{ C_{ii}^{10\dagger}C_{i'i'}^{10},\;\rho\right\} \right).\label{eq:Lindblad_RWA}
\end{alignat}
The first part $\gamma_{ij;i'j'}^{\alpha\alpha;\alpha'\alpha'}$ describes
intra-band decoherence\textbf{ }effects. The second part $\gamma_{ii;i'i'}^{01;01}$
describes polaron relaxation process from upper to lower band, while
the third part $\gamma_{ii;i'i'}^{10;10}$ describes an excitation
from lower to upper band. We use the short notation $\gamma_{i,j}^{01}\equiv\gamma_{ii;jj}^{01;01}$
and $C_{i}^{01}\equiv\left(\hat{a}_{i}^{0\dagger}\hat{a}_{i}^{1}\right)$
for the second term.

In this paper we focus on inter-band relaxation\textbf{ }processes.
The relaxation rate $\gamma_{i,j}^{01}$ is:
\begin{equation}
\gamma_{i,j}^{01}=2\cdot{\bf \text{Re}}\int_{0}^{\infty}d\tau e^{i\varepsilon_{{\rm P}}^{\Delta}\tau/\hbar}\langle B_{ii}^{01\dagger}\left(\tau\right)B_{jj}^{01}\left(0\right)\rangle_{T}\equiv2\cdot{\bf \text{Re}}\int_{0}^{\infty}d\tau e^{i\varepsilon_{{\rm P}}^{\Delta}\tau/\hbar}g_{ij}^{01}\left(\tau\right),\label{eq:decay_rate}
\end{equation}
with the correlation function $g_{ij}^{01}\left(\tau\right)\equiv g_{ii;jj}^{01;01}\left(\tau\right)$:

\begin{alignat}{1}
g_{ij}^{01}\left(\tau\right)= & \langle B_{ii}^{01\dagger}\left(\tau\right)B_{jj}^{01}\left(0\right)\rangle_{T}\nonumber \\
= & \sum_{\mathbf{q}\mathbf{\mathbf{q}'}}\hbar^{2}\omega_{\mathbf{q}}\omega_{\mathbf{\mathbf{q}'}}\left(M_{i,\mathbf{q}}^{01}\right)\left(M_{j,\mathbf{\mathbf{q}'}}^{01}\right)^{*}\langle\left[\left(\hat{K}_{i,i}^{10}\left(\tau\right)\right)\hat{b}_{\mathbf{q}}\left(\tau\right)+\hat{b}_{-\mathbf{q}}^{\dagger}\left(\tau\right)\left(\hat{K}_{i,i}^{10}\left(\tau\right)\right)+\left(\Lambda_{i,\mathbf{q}}^{0}-\Lambda_{i,\mathbf{q}}^{1}\right)^{*}\left(\hat{K}_{i,i}^{10}\left(\tau\right)-\langle\hat{K}_{i,i}^{10}\left(\tau\right)\rangle_{T}\right)\right]\nonumber \\
 & \qquad\qquad\qquad\cdot\left[\hat{b}_{\mathbf{\mathbf{q}'}}^{\dagger}\left(0\right)\left(\hat{K}_{j,j}^{01}\left(0\right)\right)+\left(\hat{K}_{j,j}^{01}\left(0\right)\right)\hat{b}_{-\mathbf{\mathbf{q}'}}\left(0\right)+\left(\Lambda_{j,\mathbf{\mathbf{q}'}}^{0}-\Lambda_{j,\mathbf{q}'}^{1}\right)\left(\hat{K}_{j,j}^{01}\left(0\right)-\langle\hat{K}_{j,j}^{01}\left(0\right)\rangle_{T}\right)\right]\rangle_{T}.\label{eq:correlation_g}
\end{alignat}
From the definition of $\hat{K}_{i,j}^{\alpha\beta}$, we have 
\begin{alignat}{1}
\hat{K}_{i,j}^{\alpha\beta}\left(t\right)\equiv & \exp\left[\sum_{\mathbf{q}}\left(\mu_{ij,\mathbf{q}}^{\alpha\beta}\right)^{*}\hat{b}_{\mathbf{q}}^{\dagger}\left(t\right)-\sum_{\mathbf{q}}\left(\mu_{ij,\mathbf{q}}^{\alpha\beta}\right)\hat{b}_{\mathbf{q}}\left(t\right)\right];\nonumber \\
\left(\mu_{ij,\mathbf{q}}^{\alpha\beta}\right)\equiv & -\left(\Lambda_{i,\mathbf{q}}^{\alpha\alpha}\right)+\left(\Lambda_{j,\mathbf{q}}^{\beta\beta}\right),\label{eq:Kab}
\end{alignat}
and the thermal averaged results from Bloch identity:
\begin{equation}
\langle\hat{K}_{i,j}^{\alpha\beta}\left(t\right)\rangle_{T}=\exp\left[-\sum_{\mathbf{q}}\left(N_{\mathbf{q}}+\frac{1}{2}\right)\left(\mu_{ij,\mathbf{q}}^{\alpha\beta}\right)^{*}\left(\mu_{ij,\mathbf{q}}^{\alpha\beta}\right)\right],
\end{equation}
as well as:
\begin{alignat}{1}
\langle\hat{K}_{i,j}^{\alpha\beta}\left(t\right)\hat{K}_{i',j'}^{\alpha'\beta'}\left(0\right)\rangle_{T}= & \exp\left[-\sum_{\mathbf{q}}\left(N_{\mathbf{q}}+\frac{1}{2}\right)\left(\mu_{ij,\mathbf{q}}^{\alpha\beta}\right)^{*}\left(\mu_{ij,\mathbf{q}}^{\alpha\beta}\right)\right]\exp\left[-\sum_{\mathbf{q}}\left(N_{\mathbf{q}}+\frac{1}{2}\right)\left(\mu_{i'j',\mathbf{q}}^{\alpha'\beta'}\right)^{*}\left(\mu_{i'j',\mathbf{q}}^{\alpha'\beta'}\right)\right]\nonumber \\
 & \quad\cdot\exp\left[-\sum_{\mathbf{q}}\left(N_{\mathbf{q}}\right)\left(\mu_{ij,\mathbf{q}}^{\alpha\beta}\right)^{*}\left(\mu_{i'j',\mathbf{q}}^{\alpha'\beta'}\right)e^{i\omega_{\mathbf{q}}t}-\sum_{\mathbf{q}}\left(N_{\mathbf{q}}+1\right)\left(\mu_{ij,\mathbf{q}}^{\alpha\beta}\right)\left(\mu_{i'j',\mathbf{q}}^{\alpha'\beta'}\right)^{*}e^{-i\omega_{\mathbf{q}}t}\right].
\end{alignat}

The elements such as $\langle\hat{b}_{\mathbf{q}}^{\dagger}\left(t\right)\hat{K}_{i,j}^{\alpha\beta}\left(t\right)\rangle_{T}$
can be obtained by using the Glauber formula $e^{\hat{A}+\hat{B}}=e^{\hat{A}}e^{\hat{B}}e^{-\frac{1}{2}[\hat{A},\hat{B}]}$
(with the condition $[\hat{A},[\hat{A},\hat{B}]]=[\hat{B},[\hat{A},\hat{B}]]=0$)
and differentiating $\hat{K}_{i,j}^{\alpha\beta}\left(t\right)$ in
Eq.~(\ref{eq:Kab}) as:
\begin{alignat}{1}
\frac{\partial}{\partial\left(\mu_{ij,\mathbf{q}}^{\alpha\beta}\right)^{*}}\hat{K}_{i,j}^{\alpha\beta}\left(t\right)= & \left[\hat{b}_{\mathbf{q}}^{\dagger}\left(t\right)-\frac{1}{2}\left(\mu_{ij,\mathbf{\mathbf{q}}}^{\alpha\beta}\right)\right]\hat{K}_{i,j}^{\alpha\beta}\left(t\right);\nonumber \\
\frac{\partial}{\partial\left(\mu_{ij,\mathbf{q}}^{\alpha\beta}\right)}\hat{K}_{i,j}^{\alpha\beta}\left(t\right)= & \hat{K}_{i,j}^{\alpha\beta}\left(t\right)\left[-\hat{b}_{\mathbf{q}}\left(t\right)-\frac{1}{2}\left(\mu_{ij,\mathbf{q}}^{\alpha\beta}\right)^{*}\right].\label{eq:b_K}
\end{alignat}
 As an example, averages of $\langle\hat{b}_{\mathbf{q}}^{\dagger}\left(t\right)\hat{K}_{i,j}^{\alpha\beta}\left(t\right)\rangle_{T}$
in Eq.~(\ref{eq:correlation_g}) can then be obtained by using Eq.~(\ref{eq:b_K})
and switching the order of differentiation and thermal averaging:
\begin{alignat}{1}
\langle\hat{b}_{\mathbf{q}}^{\dagger}\left(t\right)\hat{K}_{i,j}^{\alpha\beta}\left(t\right)\rangle_{T}= & \frac{\partial}{\partial\left(\mu_{ij,\mathbf{q}}^{\alpha\beta}\right)^{*}}\langle\hat{K}_{i,j}^{\alpha\beta}\left(t\right)\rangle_{T}+\frac{1}{2}\left(\mu_{ij,\mathbf{q}}^{\alpha\beta}\right)\langle\hat{K}_{i,j}^{\alpha\beta}\left(t\right)\rangle_{T}\nonumber \\
= & -N_{\mathbf{q}}\left(\mu_{ij,\mathbf{q}}^{\alpha\beta}\right)\exp\left[-\sum_{\mathbf{q}}\left(N_{\mathbf{q}}+\frac{1}{2}\right)\left(\mu_{ij,\mathbf{q}}^{\alpha\beta}\right)^{*}\left(\mu_{ij,\mathbf{q}}^{\alpha\beta}\right)\right].\label{eq:<b_K>}
\end{alignat}

Finally the correlation function Eq.~(\ref{eq:correlation_g}) for
inter-band relaxation is: 
\begin{alignat}{1}
g_{ij}^{01}\left(\tau\right)= & \sum_{\mathbf{q}}\left(\hbar\omega_{\mathbf{q}}|M_{\mathbf{q}}^{01}|\right)^{2}\cos\left[q_{x}(i-j)d\right]\left[\left(N_{\mathbf{q}}+1\right)e^{-i\omega_{\mathbf{q}}\tau}+\left(N_{\mathbf{q}}\right)e^{\text{+}i\omega_{\mathbf{q}}\tau}\right]\langle\hat{K}_{i,i}^{10}\left(\tau\right)\hat{K}_{j,j}^{01}\left(0\right)\rangle_{T}\nonumber \\
 & +\sum_{\mathbf{q}\mathbf{\mathbf{q}'}}\hbar^{2}\omega_{\mathbf{q}}\omega_{\mathbf{\mathbf{q}'}}\left(M_{\mathbf{q}}^{01}\right)\left(M_{\mathbf{\mathbf{q}'}}^{01}\right)\left(\Lambda_{\mathbf{q}}^{0}-\Lambda_{\mathbf{q}}^{1}\right)\left(\Lambda_{\mathbf{\mathbf{q}}'}^{0}-\Lambda_{\mathbf{q}'}^{1}\right)\sin\left[q_{x}(i-j)d\right]\sin\left[q'_{x}(i-j)d\right]\nonumber \\
 & \quad\cdot\left[\left(N_{\mathbf{q}}+1\right)e^{-i\omega_{\mathbf{q}}\tau}-\left(N_{\mathbf{q}}\right)e^{\text{+}i\omega_{\mathbf{q}}\tau}\right]\cdot\left[\left(N_{\mathbf{\mathbf{q}}'}+1\right)e^{-i\omega_{\mathbf{q}'}\tau}-\left(N_{\mathbf{q}'}\right)e^{\text{+}i\omega_{\mathbf{q}'}\tau}\right]\langle\hat{K}_{i,i}^{10}\left(\tau\right)\hat{K}_{j,j}^{01}\left(0\right)\rangle_{T}.\label{eq:correlation_g-1}
\end{alignat}
with
\begin{alignat}{1}
 & \langle\hat{K}_{i,i}^{10}\left(\tau\right)\hat{K}_{j,j}^{01}\left(0\right)\rangle_{T}\nonumber \\
= & \exp\left\{ \sum_{\mathbf{q}}\left(\Lambda_{\mathbf{q}}^{0}-\Lambda_{\mathbf{q}}^{1}\right)^{2}\left[-\left(2N_{\mathbf{q}}+1\right)+\left(N_{\mathbf{q}}\right)\cos\left[q_{x}(i-j)d\right]e^{+i\omega_{\mathbf{q}}\tau}+\left(N_{\mathbf{q}}+1\right)\cos\left[q_{x}(i-j)d\right]e^{-i\omega_{\mathbf{q}}\tau}\right]\right\} .\label{eq:KiKj}
\end{alignat}
The first term in correlation function (\ref{eq:correlation_g-1})
is a single-phonon process, which is similar to Fermi's Golden Rule
except a renormalization factor $\langle\hat{K}_{i,i}^{10}\left(\tau\right)\hat{K}_{j,j}^{01}\left(0\right)\rangle_{T}$.
The second term describes higher order process involving two phonons,
which are absent in Fermi's Golden Rule.

The corresponding relaxation rate $\gamma_{i,j}^{01}$ is calculated
by Eq.~(\ref{eq:decay_rate}). At zero temperature, this corresponds
to spontaneous relaxation process:
\begin{alignat}{1}
\gamma_{i,j}^{01}=\: & 2\cdot{\bf \text{Re}}\int_{0}^{\infty}d\tau e^{i\varepsilon_{{\rm P}}^{\Delta}\tau/\hbar}\cdot\sum_{\mathbf{q}}\left(\hbar\omega_{\mathbf{q}}|M_{\mathbf{q}}^{01}|\right)^{2}\cos\left[q_{x}(i-j)d\right]e^{\text{-}i\omega_{\mathbf{q}}\tau}\cdot\langle\hat{K}_{i,i}^{10}\left(\tau\right)\hat{K}_{j,j}^{01}\left(0\right)\rangle_{T}\nonumber \\
 & +2\cdot{\bf \text{Re}}\int_{0}^{\infty}d\tau e^{i\varepsilon_{{\rm P}}^{\Delta}\tau/\hbar}\cdot\sum_{\mathbf{q}\mathbf{\mathbf{q}'}}\hbar^{2}\omega_{\mathbf{q}}\omega_{\mathbf{\mathbf{q}'}}\left(M_{\mathbf{q}}^{01}\right)\left(M_{\mathbf{\mathbf{q}'}}^{01}\right)\left(\Lambda_{\mathbf{q}}^{0}-\Lambda_{\mathbf{q}}^{1}\right)\left(\Lambda_{\mathbf{q}'}^{0}-\Lambda_{\mathbf{q}'}^{1}\right)\nonumber \\
 & \quad\cdot\sin\left[q_{x}(i-j)d\right]\sin\left[q'_{x}(i-j)d\right]e^{-i\omega_{\mathbf{q}}\tau}e^{-i\omega_{\mathbf{q}'}\tau}\cdot\langle\hat{K}_{i,i}^{10}\left(\tau\right)\hat{K}_{j,j}^{01}\left(0\right)\rangle_{T}.\label{eq:decay_rate_1}
\end{alignat}
The second part, which involves higher order correlations with two
phonons being emitted, is neglected in the total decay rate in Eq.
(\ref{eq:gamma_polaron}).

%\end{widetext}

\twocolumngrid

\end{document}